\documentclass[apjl, iop]{emulateapj}

\slugcomment{Accepted for publication in AJ}

\usepackage[tight]{subfigure}
\usepackage[colorlinks]{hyperref}
\hypersetup{ colorlinks, linkcolor=blue, citecolor=blue, anchorcolor=blue}
\usepackage{color}

\graphicspath{{images/}}

\newcommand{\DR}{$\Delta m_{15}(B)$}

\newcommand{\kms}{km\,s$^{-1}$}
\newcommand{\SiII}{Si{\sc ii}}
\newcommand{\SII}{S~{\sc ii}}
\newcommand{\NaI}{Na~{\sc i}~D}

\newcommand{\OI}{O~{\sc i}}
\newcommand{\CII}{C~{\sc ii}}
\newcommand{\FeIII}{Fe~{\sc iii}}
\newcommand{\MgII}{Mg~{\sc ii}}

\newcommand{\FeII}{Fe~{\sc ii}}
\newcommand{\CaII}{Ca~{\sc ii}}
\newcommand{\Nifs}{$^{56}$Ni}
\newcommand{\ld}{$\lambda$}
\shorttitle{SN 2013dy}
\shortauthors{Zhai et al.}

\begin{document}

\title{UV-Optical Observation of Type Ia Supernova SN 2013\MakeLowercase{dy} in NGC 7250}

\author{Qian Zhai\altaffilmark{1,2,3}, Ju-Jia Zhang\altaffilmark{1,3}$^{\dagger}$, Xiao-Feng Wang\altaffilmark{4},Tian-Meng Zhang\altaffilmark{5},  Zheng-Wei Liu\altaffilmark{6}, Peter J. Brown\altaffilmark{7}, Fang Huang\altaffilmark{4},  Xu-Lin Zhao\altaffilmark{4}, Liang Chang\altaffilmark{1,3},  Wei-Min Yi\altaffilmark{1,3}, Chuan-Jun Wang\altaffilmark{1,3},Yu-Xin Xin\altaffilmark{1,3}, Jian-Guo Wang\altaffilmark{1,3}, Bao-Li Lun\altaffilmark{1,3}, Xi-Liang Zhang\altaffilmark{1,3}, Yu-Feng Fan\altaffilmark{1,3}, Xiang-Ming Zheng\altaffilmark{1,3} and Jin-Ming Bai\altaffilmark{1,3}}

\altaffiltext{1}{Yunnan Observatories (YNAO), Chinese Academy of Sciences, Kunming 650011, China; (jujia@ynao.ac.cn)}
\altaffiltext{2}{University of Chinese Academy of Sciences, Chinese Academy of Sciences, Beijing 100049, China}
\altaffiltext{3}{Key Laboratory for the Structure and Evolution of Celestial Objects, Chinese Academy of Sciences, Kunming 650011, China}
\altaffiltext{4}{Physics Department and Tsinghua Center for Astrophysics (THCA), Tsinghua University, Beijing 100084, China}
\altaffiltext{5}{National Astronomical Observatories of China (NAOC), Chinese Academy of Sciences, Beijing 100012, China}
\altaffiltext{6}{Argelander-Institut f\"{u}r Astronomie, Auf dem H\"{u}gel 71, D-53121, Bonn, Germany}
\altaffiltext{7}{George P. and Cynthia Woods Mitchell Institute for Fundamental Physics $\&$ Astronomy, Texas A. $\&$ M. University, Department of Physics and Astronomy, 4242 TAMU, College Station, TX 77843, USA}
\begin{abstract}
Extensive and independent observations of  Type Ia supernova (SN Ia) SN 2013dy  are presented, including a larger set of $UBVRI$ photometry and optical spectra from a few days before the peak brightness to $\sim$\,200 days after explosion, and ultraviolet (UV) photometry spanning from $t\,\approx\,-10$ days to $t\,\approx\,+15$ days referring to the $B$ band maximum.  The peak brightness (i.e., $M_{\rm B}\,=\,-19.65\,\pm\,0.40$ mag; $L_{\rm max}\,=\,[1.95\,\pm\,0.55]\,\times\,10^{43}$ erg s$^{-1}$) and the mass of synthesised $^{56}$Ni (i.e., $M$($^{56}$Ni)\,=\,0.90\,$\pm$\,0.26\,M$_{\sun}$) are calculated, and they conform to the expectation for a SN Ia with a slow decline rate (i.e., \DR\,=\,0.90\,$\pm$\,0.03 mag; \citealp{Phillip93}). However, the near infrared (NIR) brightness of this SN (i.e., $M_{\rm H}\,=\,-17.33\,\pm\,0.30$\,mag) is at least 1.0 mag fainter than usual.  Besides, spectroscopy classification reveals that SN 2013dy resides on the border of ``core normal" and ``shallow silicon" subclasses in the \citet{Branch09} classification scheme, or on the border of the ``normal velocity" SNe Ia and 91T/99aa-like events in the \citet{Wang09a} system. These suggest that SN 2013dy is a slow-declining SN Ia located on the transitional region of nominal spectroscopic subclasses and might not be a typical normal sample  of SNe Ia. 
\end{abstract}

\keywords {supernovae:general -- supernovae: individual (SN 2013dy), -- galaxies: individual (NGC 7250).}

\section{Introduction}
\label{sect:Intro}
\begin{figure}
\centering
\includegraphics[width=7.5cm,angle=0]{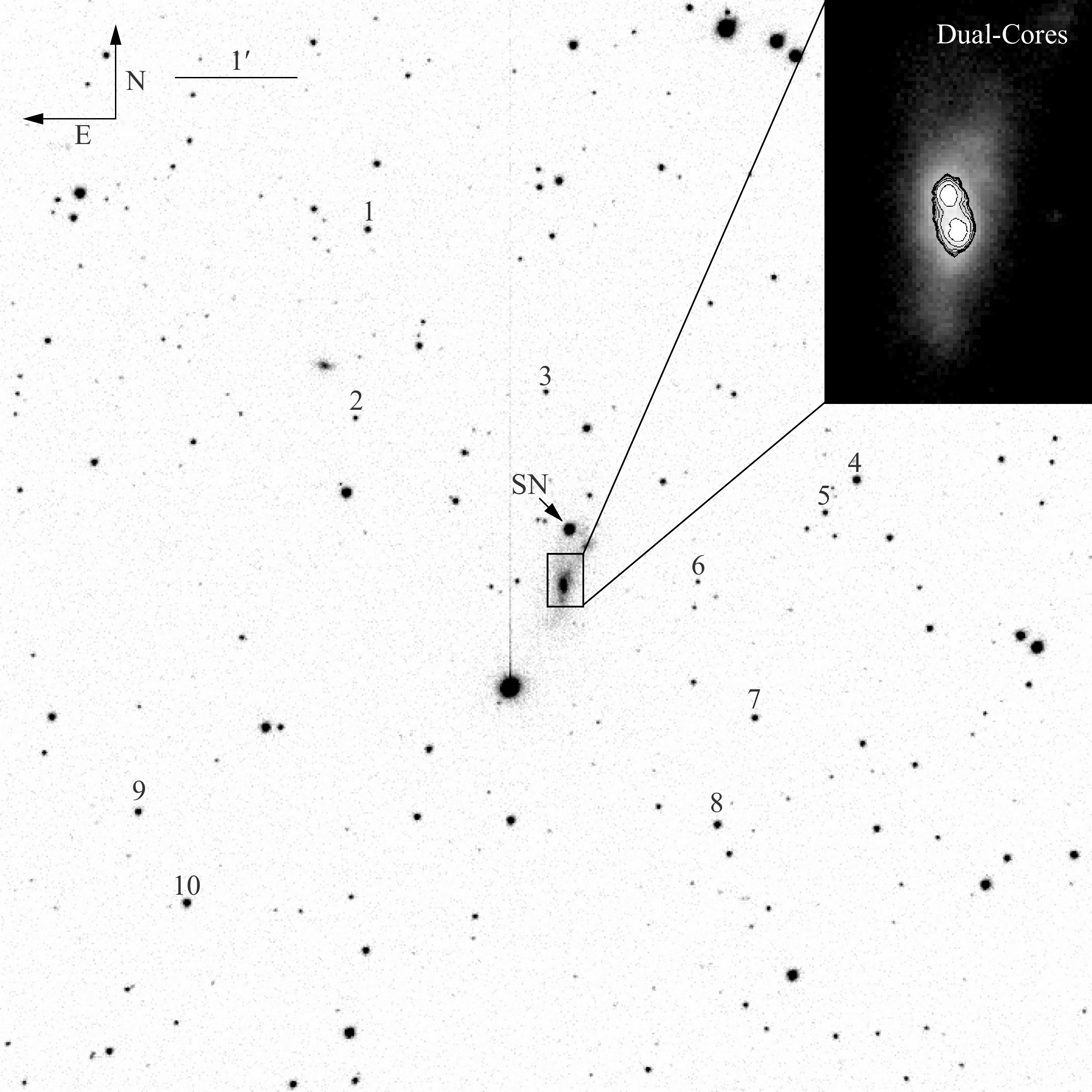}
 \caption{$R$ band image of  SN 2013dy and its reference stars, taken with the LJT and YFOSC on 2013 July 27.73.  The mean FWHM of this image is $\sim1\arcsec.10$ under the scale of $\sim0\arcsec.283$/pixel. }
\label{<img>}
\end{figure}

SN 2013dy, an SN Ia,  was discovered at roughly a magnitude of $\sim$17.2 mag on UT July 10.45  2013 (Universal Time is used throughout this paper) in an unfiltered image of the  galaxy NGC 7250  by the Lick observatory supernova search  \citep{CBET}. Its coordinates are R.A.$\,=\,22^{\rm h}\,18^{\rm m}\,17^{\rm s}$.60, Dec\,=\,$+40^\circ\,$34\arcmin\,9.54\arcsec (J2000) and it is located at 2.\arcsec3\ west and 26\arcsec.4\ north of the center of the host galaxy (see Figure \ref{<img>}). It was well studied by Zheng et al. (2013, hereafter Z13) at early phase, based on the dense photometries ($t\approx$ $-17$ to $+3$ days; the variable $t$ denotes the time since $B$ band maximum and is used throughout this paper) and spectra ($t\approx$ $-16$ to $-6$ days).  Pan et al. (2015, hereafter P15) also investigated this SN from the  $BVrRiIZYJH$ band photometry and  UV--optical spectroscopy data set spanning from $\sim$ 0.1 to $\sim$500 days after explosion.

Based on the discovery and pre-discovery images, Z13 constrained the first-light time (i.e., JD 2456483.18) of SN 2013dy to be only 2.4\,$\pm$\,1.2 hr before the first detection. This makes it the earliest known detection of an SN Ia. They inferred  an upper limit on the radius of the progenitor star of  $R_0\leq$\,0.25$R_{\sun}$ through the early-time observations, which is consistent with that of a white dwarf progenitor. The early rising light curve exhibits a broken power law with exponents of 0.88 and then 1.80, which suggests that the rising exponent of SNe Ia may vary with time. Besides, Z13 derived that SN 2013dy reached a $B$ band maximum at $\sim$17.7 days after first light with $m(B)_{\rm max}$\,=\,13.28\,$\pm$\,0.01 mag.  A spectrum taken at $t \approx -16$ days  reveals a \CII\,$\lambda$6580 absorption line comparable in strength to \SiII\,$\lambda$6355. Such a strong \CII\,lines are not usually seen in normal SNe Ia, but similar features have been observed in a few superluminous SNe Ia (i.e., SN 2009dc; \citealp{Tau09dc}). This feature suggests that the progenitor star had significant unburned material. 

In this paper, we present extensive and independent UV-optical photometry and  optical spectroscopy of SN 2013dy. Additionally, the $BVRI$ photometry of Z13 and the $BVRIYJH$ photometry of P15 are also involved in the analysis. Note that, the UV and $U$ band photometry in this paper are unique and important for the further investigation since these data are not involved in Z13 and P15. On the other hand, our dense $BVRI$ photometry and low-resolution spectra can fill the observational gaps in Z13 and P15 in the first 200 days after explosion, which makes SN 2013dy  to be a super well-sampled  SN Ia.  That could provide highly constraining information in investigating  the  properties of SNe Ia. Furthermore, the large data set  presented in this paper can help us further understand the diversity of SNe Ia and their impact on cosmological applications. 
 
The organization of this paper is as follows.  Observations and data reductions  are described in Section \ref{sect:obs}. Section \ref{sect:LC} investigates the light and color curves, and  estimates the extinction owing to the host galaxy. Section \ref{sect:Sp} presents the spectra evolution.  In Section \ref{sect:Disc}, we estimate the distance of this SN, construct the spectral energy distribution (SED) and bolometric light curve, estimate the mass of synthesized $^{56}$Ni, and discuss the spectroscopic classification. A brief summary is given in Section \ref{sect:con}.

\section{Observations and Data Reductions}
\label{sect:obs}
NGC 7250, the host galaxy of SN 2013dy,  is an irregular galaxy characterized by the ``dual cores" in the center, see the top right panel of Figure \ref{<img>}. These two cores at a distance of $\sim\,2\arcsec.38$ or $\sim\,230$ pc refer to the distance of NGC 7250 derived in this paper (i.e., $D$\,=\,20.0\,$\pm$\,4.0 Mpc). The dual-core feature might be interpreted by the interaction of two galaxies in the past.  Z13 also noted that there is a bright, blue region at  $\sim$ $8.\arcsec7$ west and $\sim\,6.\arcsec4$ south of this SN (projected distance $D$ = $\sim$\,1.2 kpc), which might relate to the merging history.

Our first observation of this supernova is in spectroscopy on 2013 July 14   (\citealp{CBET3}; 4.76 days after the first light and published in Z13) with the Yunnan Faint Object Spectrograph and Camera (YFOSC; \citealp{JJzhang14}) mounted at the Li-Jiang 2.4 m telescope (hereafter LJT; \citealp{FYF}) of Yunnan  Observatories (YNAO), China.  About two weeks later, we started  to monitor  this transient intensively at LJT in both photometry and spectroscopy spanning from $t\approx +0$ to $t\approx+180$ days.  Optical photometry data are also collected with the Tsinghua-NAOC 0.8 m telescope (hereafter  TNT; \citealp{wang08,TNT}) at Xing-Long Observation of National Astronomical Observatories (NAOC), China, from $t \approx -2$ days to $t \approx +150$ days. Additionally, three spectra were obtained at the Xing-Long 2.16 m telescope (hereafter  XLT) with Bei-Jing Faint Object Spectrograph and Camera (BFOSC).  Furthermore, this target was also observed by the UVOT \citep{UVOT05} on board the $Swift$ satellite \citep{Swift04}, spanning from $t \approx -10$ days to $t \approx +15$ days in the UV and optical bands.

\begin{deluxetable}{cccccc}
\tablecaption{ Photometric Standards in the Field of SN 2013dy}
\tablewidth{0pt}
\tablehead{\colhead{Star}&  \colhead{$U$(mag)} & \colhead{$B$(mag)} & \colhead{$V$(mag)} & \colhead{$R$(mag)} & \colhead{$I$(mag)} }
\startdata
1&	16.49(02)	&	16.32(02)	&	15.61(02)	&	15.29(01)	&	14.93(01)	\\
2&	17.61(04)	&	17.27(03)	&	16.57(01)	&	16.12(01)	&	15.78(01)	\\
3&	17.26(03)	&	17.09(03)	&	16.47(02)	&	16.05(01)	&	15.73(01)	\\
4&	16.52(02)	&	15.92(01)	&	15.09(01)	&	14.59(01)	&	14.22(01)	\\
5&	16.79(02)	&	16.72(01)	&	16.16(01)	&	15.78(01)	&	15.48(01)	\\
6&	18.18(02)	&	17.81(02)	&	17.05(03)	&	16.65(02)	&	16.26(01)	\\
7&	16.31(02)	&	16.27(03)	&	15.68(01)	&	15.34(01)	&	15.03(02)	\\
8&	15.77(02)	&	15.73(01)	&	15.18(01)	&	14.87(01)	&	14.56(01)	\\
9&	16.34(02)	&	16.20(01)	&	15.58(01)	&	15.22(01)	&	14.90(01)	\\
10&	16.36(03)	&	15.79(03)	&	14.97(01)	&	14.52(02)	&	14.18(01)	
\enddata
\tablecomments{ Uncertainties, in units of 0.01 mag, are 1 $\sigma$. 
See  Figure \ref{<img>} for the finding chart of SN 2013dy and the reference  stars.}
\label{Tab:Photo_stand}
\end{deluxetable}

\begin{figure*}
\centering
\includegraphics[width=15cm,angle=0]{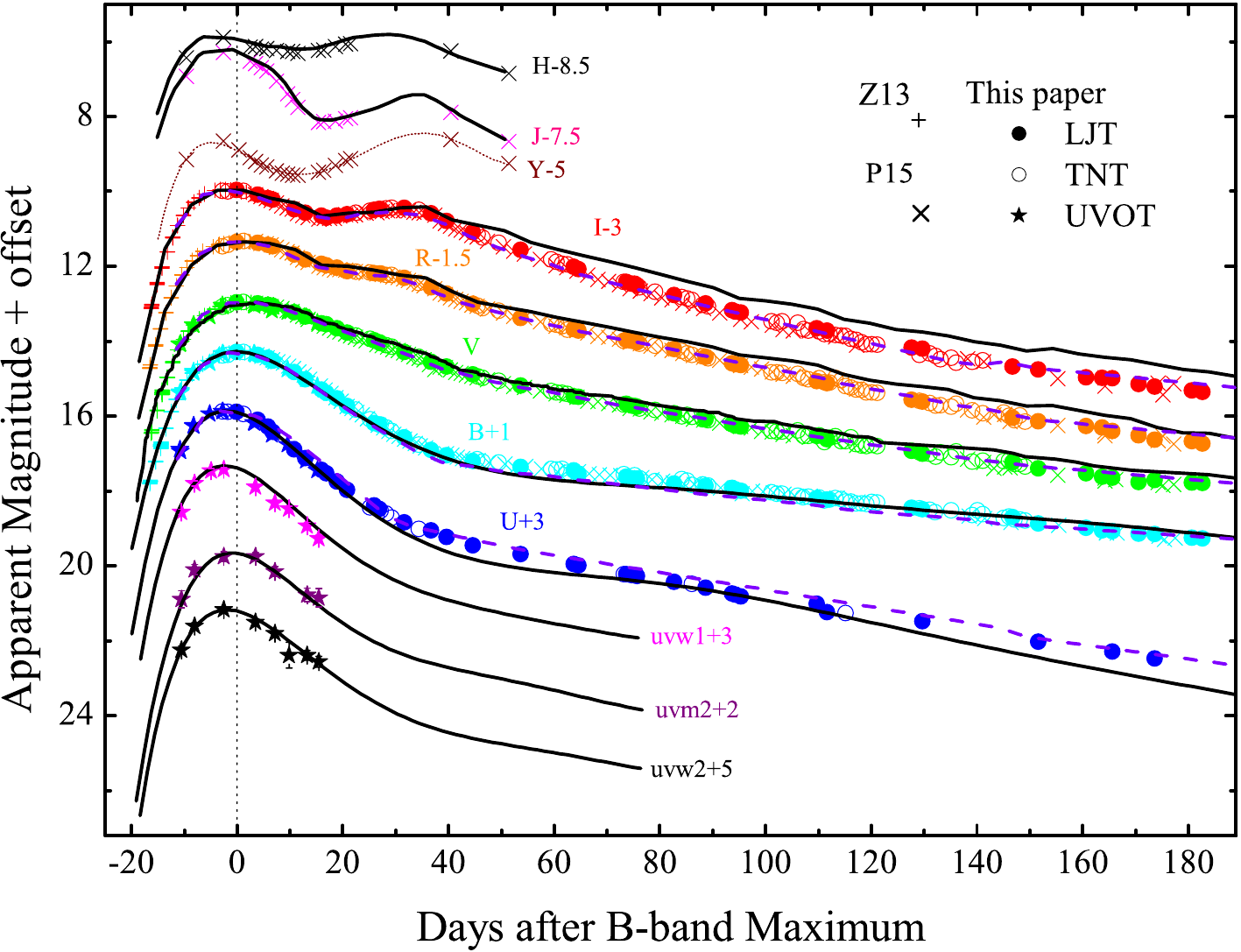}
 \caption{UV-optical-NIR light curves of SN 2013dy, which are shifted vertically for better display.  The  solid lines are derived from the photometry of SN 2011fe and the  dashed lines from that of SN 2003du. The dotted line is for the interpolation of the $Y$ band based on the photometry of SN 2013dy. See the text for detail. }
\label{<LC>}
\end{figure*}

\subsection{Photometry}
\label{subsect:Ph}

\subsubsection{Ground-based Observation}
The ground-based optical photometry of SN 2013dy were obtained in the $UBVRI$ bands with the LJT  and TNT, covering the period from $t\approx-2$ days to $t\approx+180$ days. All of the CCD images are reduced using the IRAF\footnote{IRAF, the Image Reduction and Analysis Facility, is distributed by the National Optical Astronomy Observatory, which is operated by the Association of Universities for Research in Astronomy (AURA), Inc. under cooperative agreement with the National Science Foundation (NSF).} standard procedure, including the corrections for bias, overscan, flat field, and removal of cosmic rays.  There were a few groups of templates obtained at LJT and TNT in the winter of 2014 when the SN is vanished. These templates are applied to the host subtraction of the corresponding data. Based on the subtracted images, we measured the instrumental magnitude of this SN through the aperture photometry of  the IRAF DAOPHOT package \citep{Stetson}. Ten local standard stars in the field of SN 2013dy are labeled in Figure \ref{<img>}. These reference stars are converted to the standard Johnson $UBV$ \citep{Johnson} and Kron-Cousins $RI$ \citep{Cousins} systems through transformations established by observing \citet{Landolt} standard stars during several photometric nights. The magnitudes of these stars, as listed in Table \ref{Tab:Photo_stand}, are then applied to transform the instrumental magnitudes of SN 2013dy to the standard $UBVRI$ system, as listed in Table \ref{Tab:GDPho}.

\begin{deluxetable*}{cccccccc}

\tablewidth{0pt}
\tablecaption{The $UBVRI$ Photometry of SN 2013dy from the Ground-based Observations}
\tablehead{\colhead{MJD} & \colhead{Day\tablenotemark{a}} & \colhead{$U$(mag)} & \colhead{$B$(mag)} & \colhead{$V$(mag)} & \colhead{$R$(mag)} & \colhead{$I$(mag)} & \colhead{Telescope} }

\startdata
56498.11	&	-2.77	&	12.87(03)	&	13.38(01)	&	13.08(01)	&	12.96(01)	&	12.98(01)	&	TNT	\\
56499.12	&	-1.76	&	12.88(02)	&	13.36(01)	&	13.04(01)	&	12.93(01)	&	13.00(01)	&	TNT	\\
56500.72	&	-0.16	&	12.89(03)	&	13.28(01)	&	12.96(01)	&	12.85(01)	&	12.96(01)	&	LJT	\\
56502.09	&	1.21	&	12.94(02)	&	13.28(01)	&	12.94(01)	&	12.82(01)	&	12.98(01)	&	TNT	\\
56504.70	&	3.82	&	13.09(03)	&	13.36(01)	&	12.99(01)	&	12.87(01)	&	13.09(01)	&	LJT	\\
56506.63	&	5.75	&	13.28(03)	&	13.44(01)	&	13.03(01)	&	12.93(01)	&	13.17(01)	&	LJT	\\
56507.62	&	6.74	&	13.42(04)	&	13.52(01)	&	13.05(01)	&	12.97(01)	&	13.21(01)	&	LJT\\
56511.65	&	10.77	&	13.89(02)	&	13.82(02)	&	13.22(02)	&	13.15(02)	&	13.37(03)	&	LJT	\\
56513.76	&	12.88	&	14.12(02)	&	14.01(02)	&	13.34(02)	&	13.26(03)	&	13.45(02)	&	LJT	\\
56516.76	&	15.88	&	14.43(04)	&	14.29(03)	&	13.52(03)	&	13.38(02)	&	13.54(03)	&	LJT	\\
56517.76	&	16.88	&	14.52(04)	&	14.38(01)	&	13.55(01)	&	13.42(01)	&	13.59(01)	&	LJT	\\
56519.75	&	18.87	&	14.75(02)	&	14.52(01)	&	13.65(01)	&	13.51(01)	&	13.64(01)	&	LJT	\\
56521.71	&	20.83	&	14.97(02)	&	14.73(01)	&	13.76(01)	&	13.59(01)	&	13.61(01)	&	LJT	\\
56523.34	&	22.46	&	\nodata	&	14.83(01)	&	13.85(01)	&	13.64(01)	&	13.57(01)	&	TNT	\\
56526.04	&	25.16	&	15.44(06)	&	15.08(01)	&	13.94(01)	&	13.67(01)	&	13.51(01)	&	TNT	\\
56526.78	&	25.90	&	15.38(02)	&	15.18(01)	&	13.97(01)	&	13.68(01)	&	13.49(01)	&	LJT	\\
56527.74	&	26.86	&	15.47(02)	&	15.23(01)	&	14.03(01)	&	13.69(01)	&	13.46(01)	&	LJT	\\
56528.03	&	27.15	&	15.53(04)	&	15.23(01)	&	14.05(01)	&	13.69(01)	&	13.48(01)	&	TNT	
	
\enddata
\tablecomments{Uncertainties (numbers in brackets), in units of 0.01 mag, are 1$\sigma$; MJD = JD$-2400000.5$ (Only a part of data are presented at here for the reason of typesetting and the rest part are listed at the end of this paper).}
\tablenotetext{a}{Relative to the $B$ band maximum, JD = 2456501.38.}
\label{Tab:GDPho}

\end{deluxetable*}


\subsubsection{Space-based Observation}

The $Swift$ observatory \citep{Swift04} began observing SN 2013dy on 2013 July 17.09  at about 10 days before the $B$ band maximum, and continued for approximately 26 days.  These photometric observations are performed in three UV filters ($uvw2$, $uvm2$, and $uvw1$) and three broadband optical filters ($uu$, $bb$, and $vv$).  The photometry  presented at here are reduced using the $Swift$ Optical/Ultraviolet Supernova Archive (SOUSA; \citealp{SOUSA}) reductions, including subtraction of the underlying host galaxy flux using $Swift$-UVOT observations from 2014 March and April. Table \ref{Tab:Swiftpho} lists the final UVOT UV/optical magnitudes of SN 2013dy. The results of $uvw2$ and $uvw1$ are also corrected for the `red tail' \citep{redtail} of each filter. The color-term corrections \citep{UVOTcali} have been further applied to the magnitudes of the UVOT optical filters to standard Johnson $UBV$ bands  when these data are plotted in the Figure \ref{<LC>}.  

\begin{deluxetable*}{cccccccccc}
\tablewidth{0pt}
\tabletypesize{\small}
\tablecaption{$Swift$ UVOT Photometry of SN 2013dy}
\tablehead{\colhead{MJD} & \colhead{Day\tablenotemark{a}} & \colhead{uvw2} & \colhead{uvw2$_{\rm rc}$\tablenotemark{b}} & \colhead{uvm2} & \colhead{uvw1} & \colhead{uvw1$_{\rm rc}$\tablenotemark{b}} & \colhead{$uu$} & \colhead{$bb$} & \colhead{$vv$} }
\startdata
56490.12	&	-10.76	&	17.24(10)	&	19.31	&	18.89(23)	&	15.56(07)	&	15.83	&	13.80(04)	&	14.27(04)	&	14.04(05)	\\
56492.61	&	-8.27	&	16.61(09)	&	18.39	&	18.10(14)	&	14.80(06)	&	14.96	&	13.13(03)	&	13.74(03)	&	13.58(04)	\\
56492.61	&	-8.27	&	\nodata	&	\nodata	&	\nodata	&	\nodata	&	\nodata	&	13.12(03)	&	13.72(03)	&	13.53(04)	\\
56492.67	&	-8.21	&	\nodata	&	\nodata	&	\nodata	&	\nodata	&	\nodata	&	13.09(03)	&	13.75(03)	&	13.41(04)	\\
56492.67	&	-8.21	&	\nodata	&	\nodata	&	\nodata	&	\nodata	&	\nodata	&	13.13(03)	&	13.66(03)	&	13.52(04)	\\
56495.69	&	-5.19	&	\nodata	&	\nodata	&	\nodata	&	14.46(06)	&	14.58	&	12.84(04)	&	\nodata	&	\nodata	\\
56498.12	&	-2.76	&	16.17(08)	&	18.01	&	17.75(13)	&	14.42(05)	&	14.61	&	12.78(04)	&	13.28(04)	&	13.02(04)	\\
56504.10	&	3.22	&	\nodata	&	\nodata	&	\nodata	&	\nodata	&	\nodata	&	13.17(03)	&	13.28(03)	&	13.03(04)	\\
56504.10	&	3.22	&	16.51(09)	&	19.60	&	17.74(10)	&	14.89(07)	&	15.50	&	13.07(03)	&	13.23(03)	&	12.95(03)	\\
56504.16	&	3.28	&	\nodata	&	\nodata	&	\nodata	&	\nodata	&	\nodata	&	13.12(03)	&	13.28(03)	&	13.04(04)	\\
56504.16	&	3.28	&	\nodata	&	\nodata	&	\nodata	&	\nodata	&	\nodata	&	13.05(03)	&	13.24(03)	&	13.00(03)	\\
56507.50	&	6.62	&	\nodata	&	\nodata	&	\nodata	&	\nodata	&	\nodata	&	13.44(04)	&	13.50(03)	&	13.09(04)	\\
56507.50	&	6.62	&	\nodata	&	\nodata	&	\nodata	&	\nodata	&	\nodata	&	13.50(03)	&	13.41(03)	&	13.08(03)	\\
56507.77	&	6.89	&	16.81(10)	&	20.23	&	18.15(12)	&	15.32(07)	&	16.31	&	13.44(04)	&	13.44(03)	&	13.04(04)	\\
56507.77	&	6.89	&	\nodata	&	\nodata	&	\nodata	&	\nodata	&	\nodata	&	13.40(03)	&	13.38(03)	&	13.02(03)	\\
56510.42	&	9.54	&	17.38(35)	&	22.51	&	\nodata	&	15.47(07)	&	16.46	&	13.76(04)	&	13.59(03)	&	\nodata	\\
56510.42	&	9.54	&	\nodata	&	\nodata	&	\nodata	&	\nodata	&	\nodata	&	13.72(04)	&	13.57(03)	&	\nodata	\\
56510.49	&	9.61	&	\nodata	&	\nodata	&	\nodata	&	\nodata	&	\nodata	&	13.73(04)	&	13.44(03)	&	\nodata	\\
56510.49	&	9.61	&	\nodata	&	\nodata	&	\nodata	&	\nodata	&	\nodata	&	13.74(04)	&	13.57(03)	&	\nodata	\\
56513.90	&	13.02	&	17.39(13)	&	21.47	&	18.78(18)	&	15.93(09)	&	17.47	&	14.18(05)	&	13.85(03)	&	13.43(04)	\\
56513.90	&	13.02	&	\nodata	&	\nodata	&	\nodata	&	\nodata	&	\nodata	&	14.16(04)	&	13.84(03)	&	13.31(04)	\\
56516.10	&	15.22	&	\nodata	&	\nodata	&	\nodata	&	\nodata	&	\nodata	&	14.54(06)	&	14.14(04)	&	13.54(04)	\\
56516.10	&	15.22	&	17.56(14)	&	21.33	&	18.86(24)	&	16.27(09)	&	18.15	&	14.44(04)	&	14.10(03)	&	13.48(04)	
\enddata
\tablecomments{Uncertainties (numbers in brackets), in units of 0.01 mag, are 1$\sigma$; MJD = JD$-2400000.5$.}
\tablenotetext{a}{Relative to the date of the $B$ band maximum, JD = 2456501.38.}
\tablenotetext{b}{After `Red-tail' correction \citep{redtail}.}
\label{Tab:Swiftpho}
\end{deluxetable*}

\begin{figure}
\centering
\includegraphics[width=8cm,angle=0]{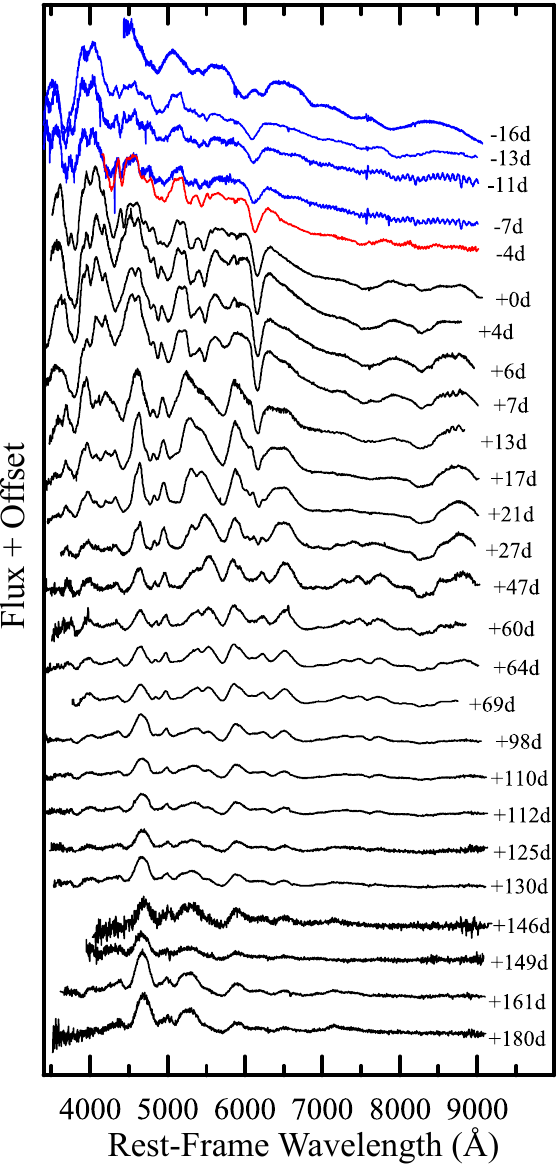}
 \caption{Optical spectra of SN 2013dy from $t\approx -16$ days to $t\approx$ +180 days with arbitrary vertical offsets for clarity. The epochs are marked for better display. Note that the spectra at $t < +0$ days are published in Z13 (blue) and P15 (red).}
\label{<Whsp>}
\end{figure}

\subsection{Spectroscopy}
\label{subsect:ObsofSP}

A journal of the spectroscopic observation of SN 2013dy is given in Table \ref{Tab:Spec_log}, containing 21 low-resolution  spectra spanning from $t\approx+0$ to $+180$ days; see also Figure \ref{<Whsp>}.  All spectra were reduced using standard IRAF long-slit spectra routines. The flux calibration was done with the standard spectrophotometric flux stars observed at a similar airmass on the same night and were double-checked with the synthetic photometry computed using \citet{Bess90} passbands. The spectra were further corrected for the atmospheric absorption and telluric lines at each observatory. 

\begin{deluxetable*}{lcccccc}
\tabletypesize{\footnotesize}
\tablewidth{0pt}
\tablecaption{Journal of Spectroscopic Observations of SN 2013dy}

\tablehead{\colhead{Date} & \colhead{MJD} & \colhead{Epoch\tablenotemark{a}} & \colhead{Res.} & \colhead{Range} & \colhead{Exp. time} & \colhead{Telescope}\\ 
\colhead{(UT)} & \colhead{(-240000.5)} & \colhead{(days)} & \colhead{(\AA)} & \colhead{(\AA)}  & \colhead{(s)} & \colhead{(+Instrument)} }

\startdata
2013 Jul. 27.71  	&	56500.71	&	-0.17	&	18	&	 3500-9100 	&	1200	&	 LJT+YFOSC	\\
2013 Jul. 31.72  	&	56504.72	&	+3.84	&	18	&	 3430-8980 	&	1200	&	 LJT+YFOSC	 \\
2013 Aug. 02.64  	&	56506.64	&	+5.76	&	18	&	 3430-8960 	&	1200	&	 LJT+YFOSC	\\
2013 Aug. 03.63  	&	56507.63	&	+6.75	&	18	&	 3410-9000 	&	1200	&	 LJT+YFOSC	 \\
2013 Aug. 09.60  	&	56513.63	&	+12.75	&	25	&	 3470-8840 	&	1800	&	 XLT+BFOSC \\
2013 Aug. 13.78  	&	56517.78	&	+16.90	&	18	&	 3430-9020 	&	1200	&	 LJT+YFOSC \\
2013 Aug. 17.71  	&	56521.71	&	+20.83	&	18	&	 3450-9010 	&	1200	&	 LJT+YFOSC\\
2013 Aug. 23.73 	&	56527.73	&	+26.85	&	18	&	 3620-8970 	&	1200	&	 LJT+YFOSC\\
2013 Sep. 12.65  	&	56547.65	&	+46.77	&	18	&	 3380-9030 	&	1200	&	 LJT+YFOSC\\
2013 Sep. 25.60  	&	56560.65	&	+59.77	&	25	&	 3500-8850 	&	2400	&	 XLT+BFOSC\\
2013 Sep. 29.50  	&	56564.50	&	+63.62	&	18	&	 3420-9010 	&	1800	&	 LJT+YFOSC \\
2013 Oct. 04.50  	&	56569.50	&	+68.62	&	25	&	 3770-8750 	&	2400	&	 XLT+BFOSC\\
2013 Nov. 02.52  	&	56598.52	&	+97.64	&	18	&	 3400-9050 	&	1800	&	 LJT+YFOSC \\
2013 Nov. 14.50  	&	56610.50	&	+109.62	&	18	&	 3400-9110 	&	1800	&	 LJT+YFOSC\\
2013 Nov. 16.49  	&	56612.49	&	+111.61	&	18	&	 3390-9130 	&	1800	&	 LJT+YFOSC\\
2013 Nov. 29.49  	&	56625.49	&	+124.61	&	18	&	 3480-9130 	&	1350	&	 LJT+YFOSC\\
2013 Dec. 03.50   	&	56630.50	&	+129.62	&	18	&	 3530-9110 	&	2700	&	 LJT+YFOSC\\
2013 Dec. 20.60  	&	56646.60	&	+145.72	&	18	&	 4070-9100 	&	3000	&	 LJT+YFOSC\\
2013 Dec. 23.54  	&	56649.54	&	+148.66	&	18	&	 3950-9080 	&	3600	&	 LJT+YFOSC\\
2014 Jan. 04.48  	&	56661.48	&	+160.60	&	18	&	 3610-9090 	&	4200	&	 LJT+YFOSC \\
2014 Jan. 23.50  	&	56680.50	&	+179.62	&	18	&	 3520-9110 	&	2700	&	 LJT+YFOSC
\enddata
\tablecomments{Journal of spectroscopic observations of SN 2013dy.}
\tablenotetext{a}{Relative to the $B$ band maximum on JD. 2456501.38.}
\label{Tab:Spec_log}
\end{deluxetable*}

\section{Light Curves of SN 2013dy}
\label{sect:LC}
Figure \ref{<LC>} shows the optical and UV light curves of SN 2013dy, overplotted with that of two well-sampled normal SNe Ia: SN 2011fe and SN 2003du. The $BVRI$ photometry of Z13 and $BVRIYJH$ photometry of P15 are also exhibited. Note that, the solid lines in this figure are based on the photometry of SN 2011fe: the UV curves are derived from the $Swift$ photometry presented in \citet{Brown11fe} and fitted with a low order polynomial; the optical curves are derived from the photometry presented in \citet{KCzhang15}; the near infrared (NIR) curves are based on the photometry presented in \citet{Matheson12}. The light curves of SN 2011fe in the UV and optical are  stretched by a  factor of 1.20 on the horizontal axis. On the other hand, the $J$ and $H$ band curves of SN 2011fe are stretched by a factor of 1.10.  The  dashed lines are based on the photometry of SN 2003du \citep{03du} with a stretch factor of 1.05.   The $Y$ band photometry of SN 2013dy (P15) are also plotted in this figure and overplotted with a polynomial fit (in dotted line).  Detailed analyses are presented in the following sections.  
 
\begin{figure*}
\centering
\includegraphics[width=16cm,angle=0]{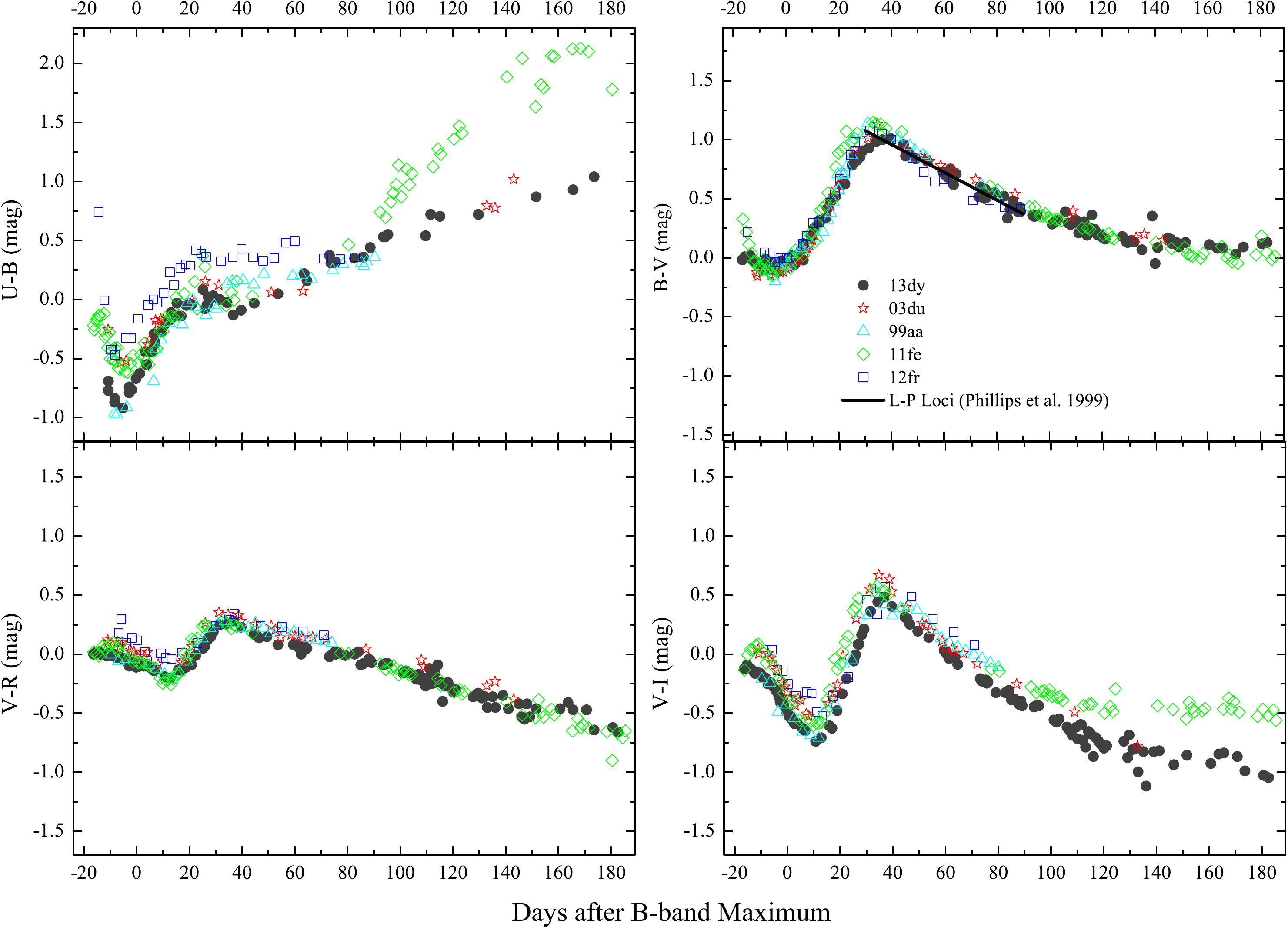}
\caption{Optical color curves of SN 2013dy compared with those of SN 1999aa, SN 2003du, SN 2011fe, and SN 2012fr, see the text for details.  }
\label{<CC>}
\end{figure*}

 \begin{deluxetable}{cccccc}
\tablewidth{0pt}
\tablecaption{ UV and Optical Light Curve Parameters of SN 2013dy}

\tablehead{\colhead{Band} & \colhead{$\lambda_{\rm eff}$} & \colhead{$t_{\rm max}$\tablenotemark{a}} & \colhead{m$_{\rm peak}$} & \colhead{$\Delta$m$_{15}$}& \colhead{M$_{\rm peak}$}  \\
\colhead{} & \colhead{(\AA)} & \colhead{} & \colhead{(mag\tablenotemark{b})} & \colhead{(mag\tablenotemark{b})} & \colhead{(mag\tablenotemark{b})}  }

\startdata
$uvw2$\tablenotemark{c} & 1928 & 498.07(40) & 17.62(10) & 2.49(20)&-16.90(60)\\
$uvm2$ & 2246 & 499.74(60) & 17.65(08) & 1.47(12) &-16.34(50)\\
$uvw1$\tablenotemark{c} & 2600 & 498.39(50) & 14.47(10) & 2.50(20) &-19.14(60)\\
$U$ & 3650 & 498.98(30) & 12.88(05) & 1.35(08)&-20.34(45) \\
$B$ & 4450 & 500.88(30) & 13.29(01) & 0.90(03) &-19.65(40)\\
$V$ & 5500 & 501.87(30) & 12.94(01) & 0.57(03) &-19.62(40)\\
$R$ & 6450 &501.45(30) & 12.83(02) & 0.54(03) &-19.62(40)\\
$I$ & 7870 & 498.72(30) & 12.95(02) & 0.47(03) &-19.54(40) \\
$Y$\tablenotemark{d} & 9100 & 496.00(40) & 13.58(04) & 0.97(05) &-18.42(35) \\
$J$\tablenotemark{d}& 12500 & 499.37(30) & 13.80(03) & 1.69(05) &-18.02(35) \\
$H$\tablenotemark{d} & 16000 & 496.41(30) & 14.37(03) & 0.39(05) &-17.33(30) 
\enddata
\tablenotetext{a}{Uncertainties of peak-light dates, in units of 0.01 day, are 1 $\sigma$. The date is MJD$-$56000.}
\tablenotetext{b}{Uncertainties of magnitudes, in units of 0.01 mag, are 1 $\sigma$. }
\tablenotetext{c}{The estimations of uvw2 and uvw1  are corrected for the `red-tail' of each filter \citep{redtail}.}
\tablenotetext{d}{Based on the NIR photometry published in P15.}
\label{Tab:LV_par}
\end{deluxetable}

\subsection{Parameters of Photometry}
\label{subsect:LC}

Based on the photometry published in Z13, P15, and this paper, we derived the parameters of peak magnitudes, maximum dates, and light curve decline rates (i.e., $\Delta m_{15}$) through a low order polynomial fit, as listed in Table \ref{Tab:LV_par}.  It is found that SN 2013dy reaches a $B$ band maximum brightness of $13.29\pm0.01$ mag on JD 2456501.38\,$\pm$\,0.30 (2013 July 27.88), which is close to that in Z13 (i.e., $m_{\rm max}(B)$\,=\,13.28 mag, JD\,=\,2456500.88) and P15 (i.e., $m_{\rm max}(B)$\,=\,13.23 mag, JD\,=\,2456501.61) roughly. The observed $B$ band decline rate  is estimated as \DR\ = 0.90\,$\pm$\,0.03 mag, which is close to the estimation in P15 (i.e., 0.92). 

One can see that in Figure \ref{<LC>}, the light curves of SN 2013dy resemble the stretched light curves of  SN 2003du (i.e., \DR\,=\,1.02; \citealp{03du}) and SN 2011fe (i.e., \DR\,=\,1.10; \citealp{Munari13}), especial at $t < +20$ days.  Note that the stretched curves of SN 2011fe decline with slower rates in the  $VRI$ bands but with  a quicker rate in  the $U$ band than that of SN 2013dy and SN 2003du.

\subsection{Color Curves}
\label{subsect:CC}
Figure \ref{<CC>} shows the optical color curves of SN 2013dy, corrected for the Galactic reddening $E(B-V)$ = 0.15\,$\pm\,0.02$ mag \citep{Schle98,Schlafly11} and the host galaxy reddening derived in Section \ref{subsect:ext}. The Galactic reddening law (i.e., $R_{\rm V}$ = 3.1; \citealp{Cardelli}) is adopted for this correction. Overplotted are the color curves of SN 1999aa \citep{Jha}, SN 2003du \citep{03du}, SN 2011fe \citep{KCzhang15}, and SN 2012fr \citep{JJzhang14}. The overall color evolution of SN 2013dy is similar to that of SN 2003du, but distinctions are present in the colors of  $U-B$ at the early phase. On the contrary, the $U-B$ color of  SN 2013dy is similar to SN 1999aa; both are bluer than the others at $t < +5$ days. It might imply a similar temperature between SN 2013dy and SN 1999aa at this period. A larger scatter appears in the $V-I$ color of SN 2013dy and SN 2011fe  at $t > +80$ days.

\begin{figure}
\centering
\includegraphics[width=8.5cm,angle=0]{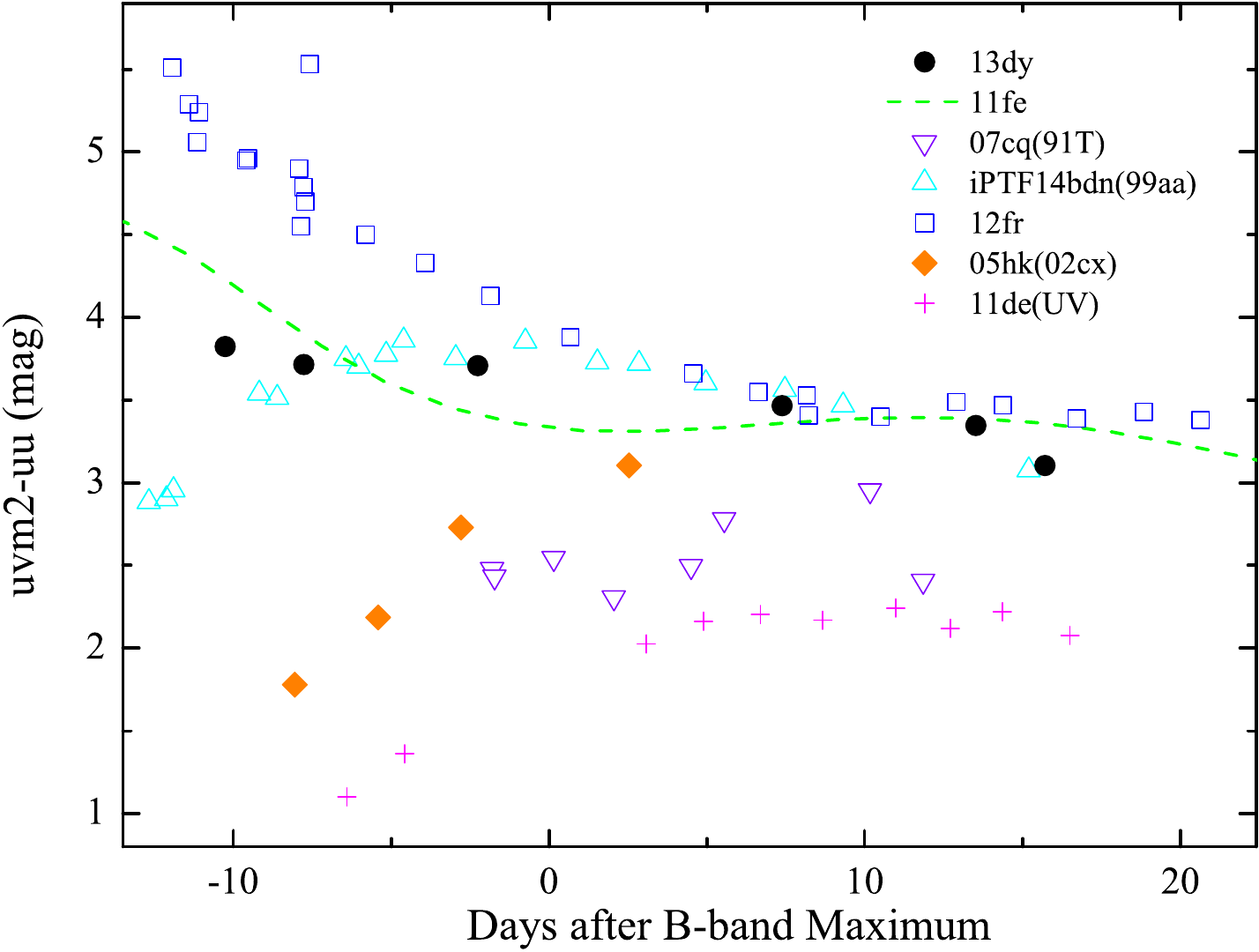}
 \caption{The $uvm2-uu$ color curve of SN 2013dy compared with that of  SN 2011fe (normal),  SN 2012fr (narrow-lined), SN 2007cq (91T-like), iPTF 14bdn (99aa-like), SN 2005hk (02cx-like) and SN 2011de (UV excess); see  the text for details. }
\label{<UVOP>}
\end{figure}

Figure \ref{<UVOP>} displays the $uvm2-uu$ color of SN 2013dy based on the observations of $Swift$-UVOT and corrected for the extinction. Comparisons are made between  the normal SN 2011fe \citep{Brown11fe}, narrow-lined SN 2012fr \citep{JJzhang14}, 91T-like \citep{Filip92a,Paolo95} event SN 2007cq \citep{SOUSA}, 99aa-like \citep{99aa} event iPTF 14bdn \citep{Smitka15},  02cx-like \citep{Li02cx} event  SN 2005hk \citep{SOUSA}, and  UV excess event SN 2011de \citep{2011de}.  These samples seem to be divided into two groups, where SN 2013dy, SN 2011fe, SN 2012fr, and iPTF 14bdn are located in the upper region with a redder color (i.e., $\geq$ 3 mag) than the remaining peculiar events (i.e., SN 2007cq, SN 2005hk, and SN 2011de).  In general, the color of SN 2013dy is similar to that of iPTF 14bdn at $t > -5$ days.  Before this period, iPTF 14bdn turns to red monotonically, which is distinct from that of SN 2011fe and SN 2012fr. On the other hand, SN 2013dy presents a flatter curve and is located in the middle of this upper group. This might relate to the transitional  position of SN 2013dy in the classification scheme (e.g., locates on the border of  normal and 91T/99aa-like events), as discussed in Section \ref{subsect:diver}.

\begin{deluxetable}{ccc}
\tablewidth{0pt}
\tablecaption{The host extinction of SN\,2013dy}

\tablehead{\colhead{Method} & \colhead{Details} & \colhead{Results (mag)} }
\startdata
EW(\NaI)\tablenotemark{a}  & 0.16EW-0.01\tablenotemark{b} & 0.07$\pm$0.05\\
EW(\NaI)\tablenotemark{a}  & 0.51EW-0.04\tablenotemark{b} & 0.23$\pm$0.05\\
EW(\NaI)\tablenotemark{a}  & 0.43EW-0.08\tablenotemark{c} & 0.15$\pm$0.05\\
Color curve  & $B_{\rm max} - V_{\rm max}$\tablenotemark{d} & 0.32$\pm$0.05 \\
Color curve  &Lira-Phillips\tablenotemark{e}& 0.20$\pm$0.05
\enddata
\tablenotetext{a}{EW(\NaI) = 0.53\AA, estimated by Z13}
\tablenotetext{b}{\citet{Turatto03}}
\tablenotetext{c}{\citet{Poznanski11}}
\tablenotetext{d}{\citet{Wang09b}}
\tablenotetext{e}{\citet{Phillip99}}
\label{Tab:extin}
\end{deluxetable}

\subsection{Extinction}
\label{subsect:ext}
The reddening due to the host galaxy can be estimated using several empirical methods. For example, the spectra published in Z13 exhibit significant \NaI\,absorption from both the host galaxy and the Milky Way. On the other hand, we can derived it according to the Lira-Phillips relation  based on the intrinsic $B-V$ color at $+30 < t < +90$ days \citep{Phillip99}. Additionally, the maximum-light color $B_{\rm max} - V_{\rm max}$ related to the decline rate can also be used to estimate the reddening of SNe Ia also (e.g.,  \citealp{Phillip99,Wang09a}). All of these methods are introduced to calculate the host galaxy reddening of SN 2013dy, as listed in Table \ref{Tab:extin}, and yield an average  value $E(B-V)_{\rm host}\,=\,0.20\,\pm\,0.10$ mag. Considering the   Galactic reddening ($E(B-V)$=0.15\,$\pm\,0.02$ mag), $E(B-V)_{\rm total}\,=\,0.35\,\pm\,0.10$ mag is thus adopted in this paper.

\section{Spectra}
\label{sect:Sp}

\subsection{Temporal Evolution}
\label{subsect:Temp}

\subsubsection{Pre-maximum}
\label{subsect:ear_sp}
\begin{figure}
\centering
\includegraphics[width=8.5cm,angle=0]{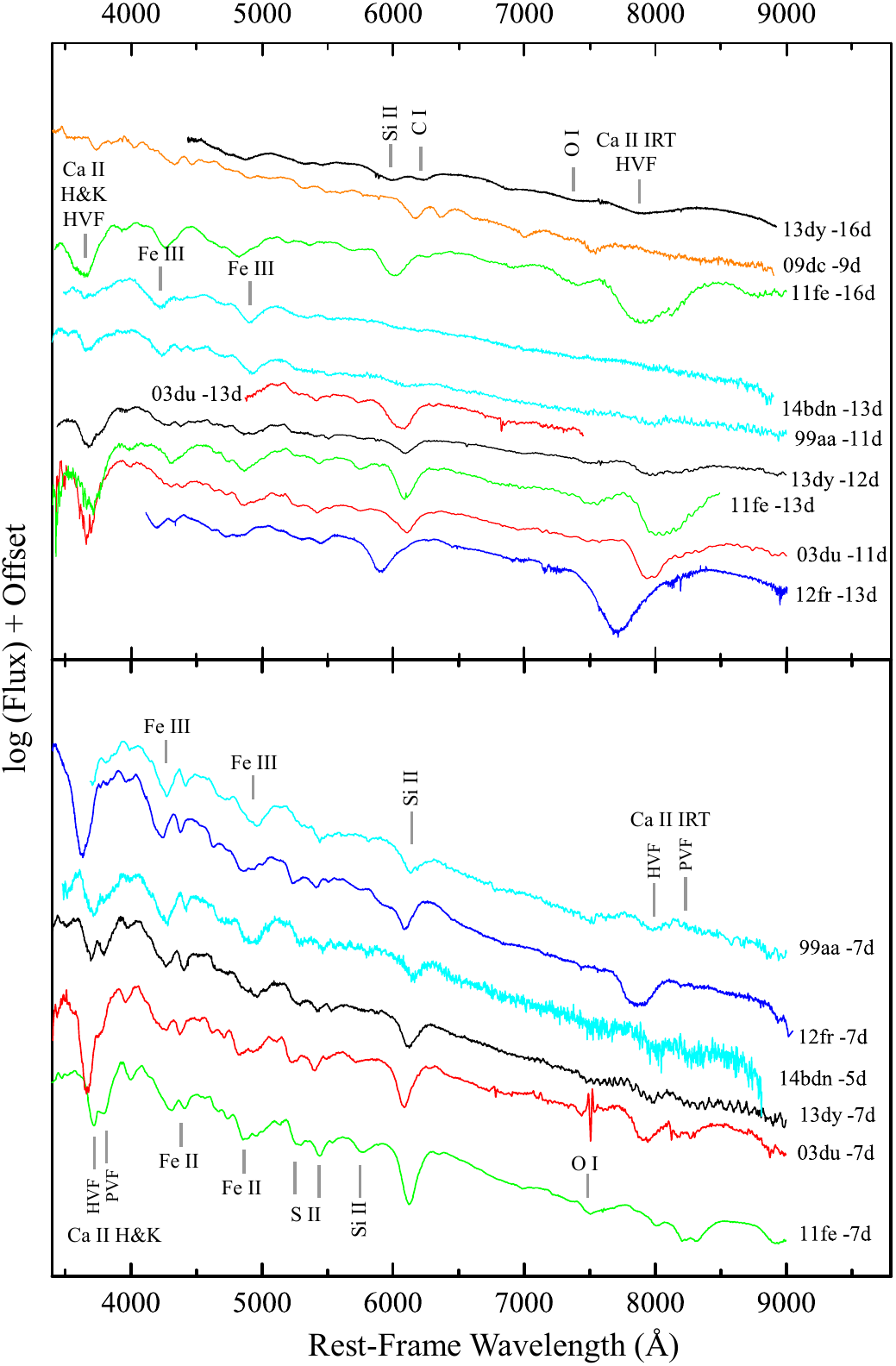}
 \caption{Early spectra of SN 2013dy overplotted with the spectra of SN 1999aa, SN 2003du, SN 2009dc, SN 2011fe, SN 2012fr and iPTF 14bdn at the selected phases. All of these spectra were corrected with redshift and reddening. }
\label{<Sp_ear>}
\end{figure}

Figure \ref{<Sp_ear>} displays the selected early spectra of SN 2013dy (Z13). Overplotted are the early spectra of SN 1999aa \citep{99aa}, SN 2003du \citep{03du}, SN 2009dc \citep{Tau09dc}, SN 2011fe \citep{Nug11fe,KCzhang15}, SN 2012fr \citep{C13,JJzhang14}, and iPTF 14bdn \citep{Smitka15}  at the similiar phases. 

The early spectra of SN 2013dy consist of absorption features from singly ionized intermediate-mass elements (IMEs, e.g., Si, S, Mg, and Ca) and  Fe, which are usually seen in normal SNe Ia. Note that the absence of \FeIII\ features in the early spectra of SN 2013dy would indicate a lower temperature than 91T/99aa events, which are characterized by  strong absorptions of double-ionized iron \citep{Paolo95}. 

The first spectrum of SN 2013dy is characterized by the strong absorptions of unburnt carbon of \CII\,$\lambda$6580. A weaker \CII\,$\lambda$7234 feature is also visible.  Though the absorption of \CII\,$\lambda$6580 is not rarely seen in the normal SNe Ia (e.g., 25\%, \citealp{Parrent11,SilverCII}), it is usually not strong. Besides SN 2013dy, such a strong absorption of  \CII\,$\lambda$6580 was only found in the superluminous Ia SN 2009dc \citep{Tau09dc}.  The spectrum of the very young SN 2011fe at  $t\approx -16$ days exhibits rather weaker \CII\,than that of SN 2013dy. On the other hand, the \CII\,$\lambda$6580 line in SN 2013dy weakens rapidly, and most of the other SNe Ia are not obtained as early as this SN. In fact,  three days later, the strength of the \CII\,$\lambda$6580 line in SN 2013dy becomes as weak as that of SN 2003du at the same phase. 

\begin{figure}
\centering
\includegraphics[width=8.5cm,angle=0]{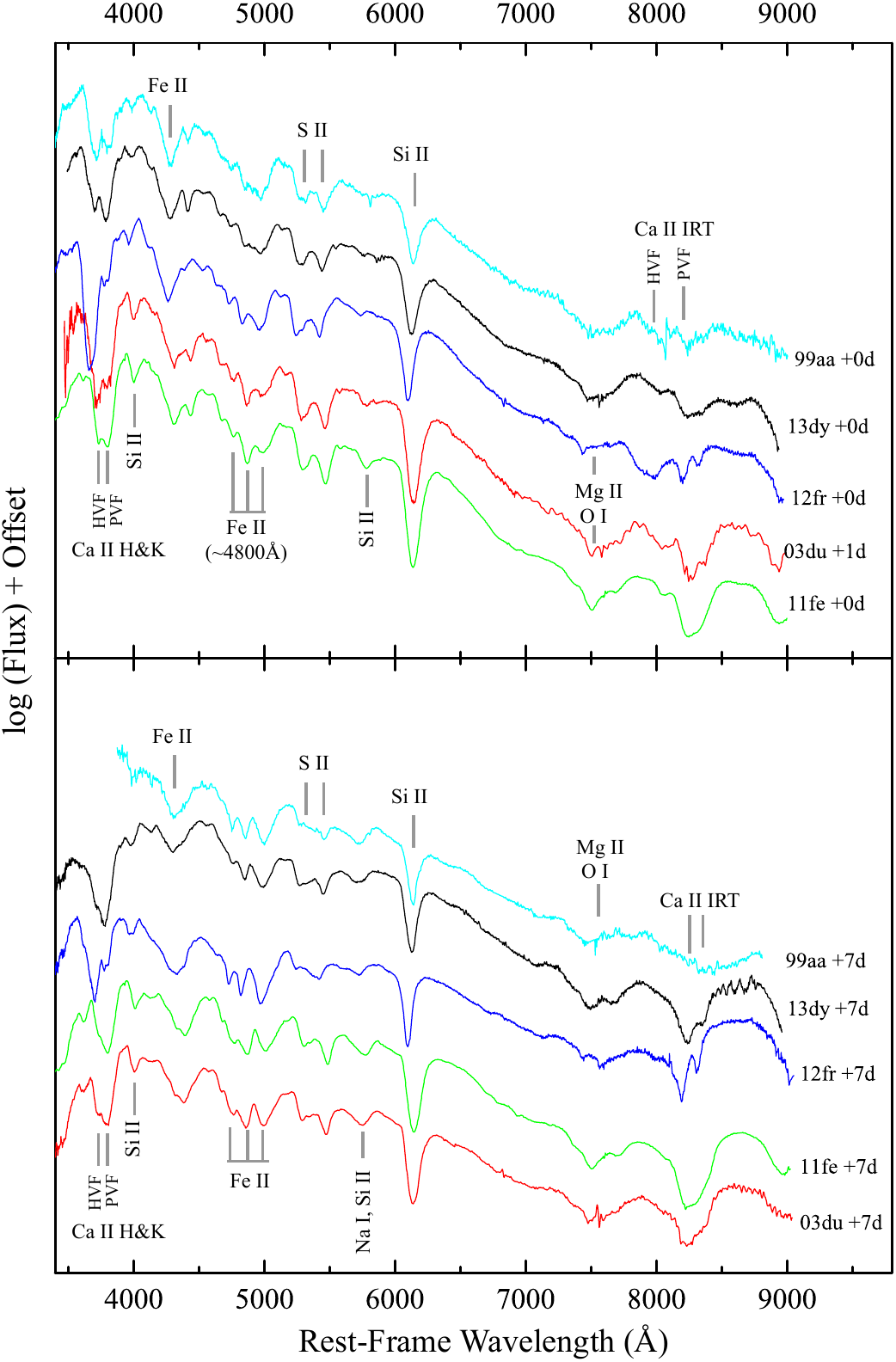}
 \caption{Spectra of SN 2013dy, SN 2012fr, SN 2011fe, SN 2003du and SN 1999aa around maximum. All of these spectra were corrected with redshift and reddening. }
\label{<Spmax>}
\end{figure}

Note that  the equivalent width (EW) ratio of \CII\,$\lambda$6580 and \SiII\,$\lambda$6355 in the spectrum of SN 2013dy at $t\,\approx\,-16$ days is $R$(C/Si)\,=\,0.48\,$\pm$\,0.04 which is similar to that of SN 2009dc at $t\,\approx\,-10$ days (i.e., $\sim$\,0.50). It might imply a similar abundance of unburnt carbon in these two SNe Ia.  However,  the velocity distribution of \CII\,$\lambda$6580 in SN 2013dy is 5700 $\pm$ 200 km s$^{-1}$ and larger than that of SN 2009dc (i.e., $\sim$ 4000 km s$^{-1}$), which might indicate a wider distribution of unburnt carbon in the outer ejecta of the former.  

At this phase, the blue-side absorption feature, so-called as high velocity features (HVFs; e.g., \citealp{Mazzali05a,Mazzali05}) are  gradually weakened in \CaII\ (i.e., H$\&$K and IR triplet)  with the emergence of the photometric component. The HVF of \CaII\,IR triplet in the spectra of SN 2013dy  begins  at $v\approx 26,000$ km s$^{-1}$.  Such an HVF is similarly seen in SN 2003du, SN 2011fe, and SN 2012fr. There is no evidence for the detached-HVF component of \SiII\,$\lambda$6355 in the spectrum of SN 2013dy at $t\approx-16$ days. Two days later, the profile of \SiII\,$\lambda$6355 becomes non-Gaussian, which might indicate two departing components in this absorption, and the bluer one should be the HVF of \SiII.

At around $t \approx -7$ days, the spectrum of SN 2013dy resembles to that of the 99aa-like event iPTF 14bdn, especially for the profile of \CaII\, H$\&$K and IR triplets, which are weaker than that in SN 2003du, SN 2011fe, and SN 2012fr. Additionally, the strength of \SiII\,\ld6355 in the spectrum of SN 2013dy is also weaker than it in the SN 2003du and SN 2011fe.  It might indicate that the  temperatures of SN 2013dy is higher than that of SN 2003du and SN 2011fe, since a higher temperature might reduce the abundance of IMEs and reproduce the observed weakening of the IME lines in objects \citep{Paolo95}.

\subsubsection{Around Maximum}
\label{subsect:sp_max}
Figure \ref{<Spmax>} displays the spectra of SN 2013dy in the first week after the $B$ band maximum compared with that of SN 1999aa \citep{99aa}, SN 2003du \citep{03du}, SN 2011fe \citep{Pereira11fe}, and SN 2012fr \citep{JJzhang14}.

\begin{figure}
\centering
\includegraphics[width=8.5cm,angle=0]{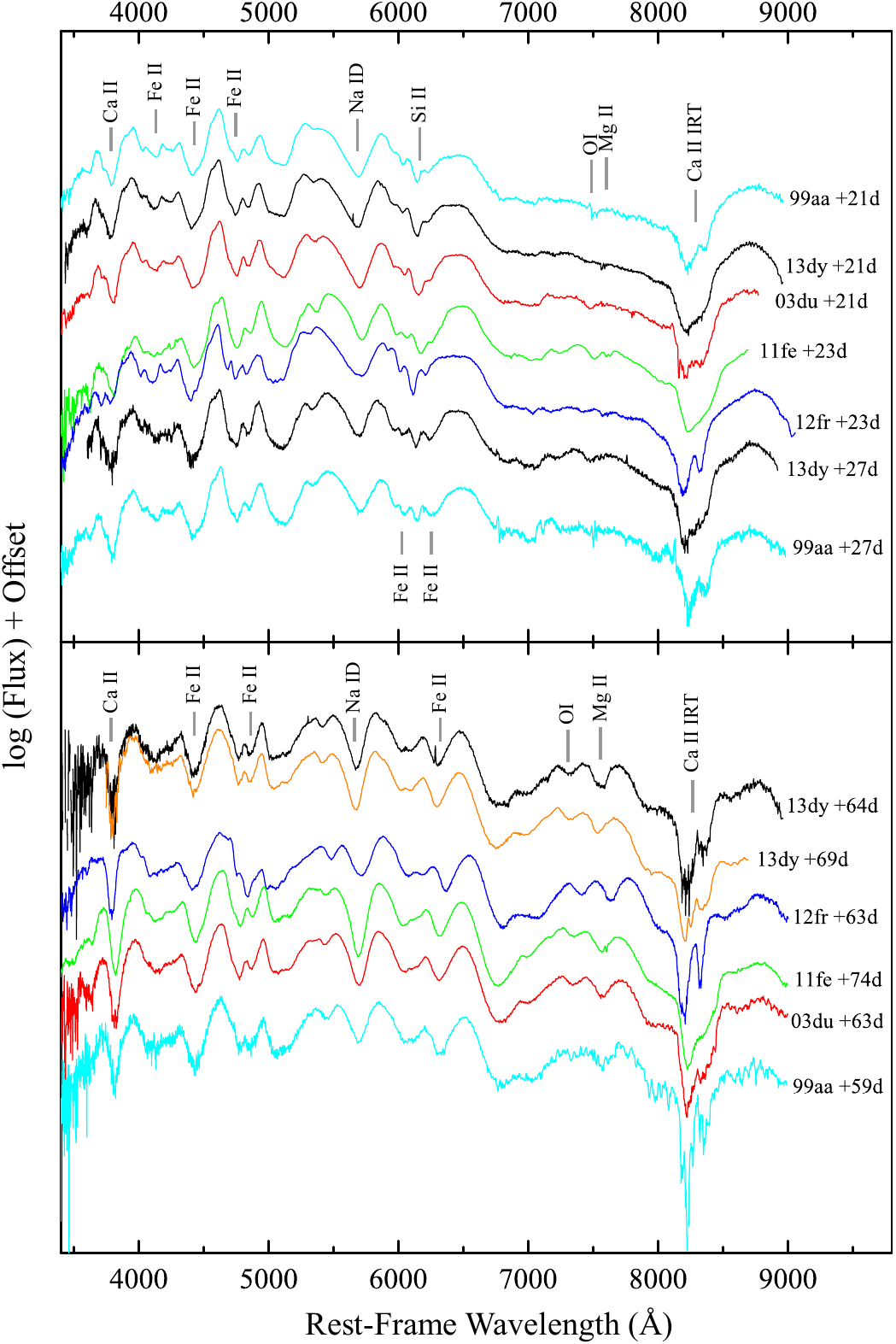}
 \caption{Spectra of SN 2013dy, SN 1999aa, SN 2003du, SN 2011fe and SN 2012fr at a few months  after maximum. All of these spectra were corrected with redshift and reddening.}
\label{<Sp_2m>}
\end{figure}

In general, the spectra of SN 2013dy resemble that of SN 1999aa at $t \approx$ +0 days. At this phase,  the absorption feature of \SiII\,$\lambda$6355 evolves to be the dominant feature in normal profile.  A minor absorption on the blue side of \NaI\,is likely due to \SiII\,$\lambda$5972, which is absent in the early spectra.  The line-strength ratio of \SiII\,$\lambda$5972 to \SiII\,$\lambda$6355, known as $R$(\SiII) \citep{Nug97}, is an approximated indicator of the photospheric temperature, with a larger value corresponding to a lower temperature and smaller \DR\  \citep{Hachinger06}. 

The ratio between \SiII\,\ld5972 and \FeII\ near $\sim$4800\AA\ (marked as $R$[Si/Fe]) shows a constant relation to the \DR\ (e.g., \citealp{Hachinger06}). \citet{Hachinger06} also suggested that the ratios of EW(\SiII\,\ld6355) and EW(\SII, known as the `w' feature) to EW(\FeII), marked as $R$(Si/Fe) and $R$(S/Fe) respectively, might indicate the intrinsic temperature and brightness of SNe Ia. The behavior for the \SII\ could be explained as the effect of increasing temperature, and the ratio between Si and Fe might reflect an abundance change. The SNe Ia with the smallest decline rates (e.g., \DR $\le$ 1.0 mag) have more Fe near the maximum-light photosphere ($\sim$ 10,000 \kms) and the intermediate decliners have more IME and less Fe at a similar velocity \citep{Mazzali07}. We note that the explosion velocity of SN 2013dy, SN 1999aa, SN 2003du, and SN 2011fe are similar and close to $\sim$ 10,000 \kms at around peak brightness. Thus, we can compare the temperature, brightness, and the mass of Fe in these SNe Ia through the $R$(S/Fe) and $R$(Si/Fe). The $R$(Si), $R$(S/Fe), $R$[Si/Fe] and $R$(Si/Fe) at $t\,\approx\,+0$ days for the selected sample are measured and listed in Table \ref{Tab:EWR}. We find that the relations of \DR\ and the EW ratios of our sample conform to the research of  \citet{Hachinger06} generally. The smaller $R$(\SiII) and  $R$(S/Fe) of SN 2013dy tend to indicate a higher temperature for this SN than that of the comparisons at $t \approx +0$ days. Moreover,  this table might suggest a similar  mass of Fe and brightness in SN 2013dy and SN 1999aa, which are larger than those in SN 2003du and SN 2011fe.

\begin{deluxetable}{lccccc}

\tablecaption{Parameters for the \DR\ -- EW ratio relation}

\tablehead{\colhead{SN} &$\Delta m_{15}$& \colhead{$R$(\SiII)\tablenotemark{a}} & \colhead{R(S/Fe)\tablenotemark{b}} & \colhead{R[Si/Fe]\tablenotemark{c}} & \colhead{R(Si/Fe)\tablenotemark{d}} } 

\startdata
2013dy&0.90 &0.05 & 0.46 & 0.02 & 0.48 \\
1999aa&0.83&0.07 & 0.55 & 0.04 & 0.58 \\
2012fr&0.85&0.07 & ... & 0.03 & ... \\
2003du&1.02& 0.11 & 0.64 & 0.08 & 0.69 \\
2011fe&1.10 &0.14 & 0.68 & 0.11 & 0.74 
\enddata
\tablenotetext{a}{EW(\SiII\,\ld5972) / EW(\SiII\,\ld6355)}
\tablenotetext{b}{EW(\SII\,`w') / EW(\FeII\,$\sim$\,4800\AA)}
\tablenotetext{c}{EW(\SII\,\ld5972) / EW(\FeII\,$\sim$\,4800\AA)}
\tablenotetext{d}{EW(\SII\,\ld6355) / EW(\FeII\,$\sim$\,4800\AA)}
\label{Tab:EWR}

\end{deluxetable}

 A notable distinction among SN 2013dy and the comparison SNe is the profile of the absorptions around 3800\AA, which can be attributed to the absorptions of \CaII\,H$\&$K. Such a difference might relate to the scatters in the $U-B$ color, and  the behavior of \CaII\,H$\&$K  at maximum light may be an indicator of intrinsic SN Ia color \citep{Chotard11,foley11,Blondin12,foley12}. Based on these samples, we find that the SNe Ia with stronger HVFs of \CaII\,H$\&$K are usually redder in $U-B$ color at around the peak brightness.

\subsubsection{A Few Months After Maximum}
\label{subsect:Sp_OneM}

At $t \sim +3$ weeks (see the upper panel of Figure \ref{<Sp_2m>}), the absorption of \SiII\,$\lambda$6355 of all the comparison SNe are contaminated by the surrounding absorption of iron-group elements. The spectra features decrease from the maximum and are dominated by the absorption of \NaI.  Around $\sim+2$ to  $\sim +3$ months after maximum, these spectra become stable without obvious evolution,  while the iron-group elements become dominant.  One interesting feature is the  \CaII\,IR triplet feature at $t\approx+69$ days, where this triplet is clearly split into three components.

Note that the velocities of  \CaII\,are similar to those of SN 2003du and SN 2011fe but slower than that of SN 2012fr. However, the absorptions of \NaI, \OI, and \MgII\,appear to be faster in the former three SNe Ia than in the latter. This might suggest that the ejecta of SN 2013dy, SN 2003du, and SN 2011fe are well mixed, and  a stratification structure might exist  in the ejecta of SN 2012fr.

\begin{figure}
\centering
\includegraphics[width=8.5cm,angle=0]{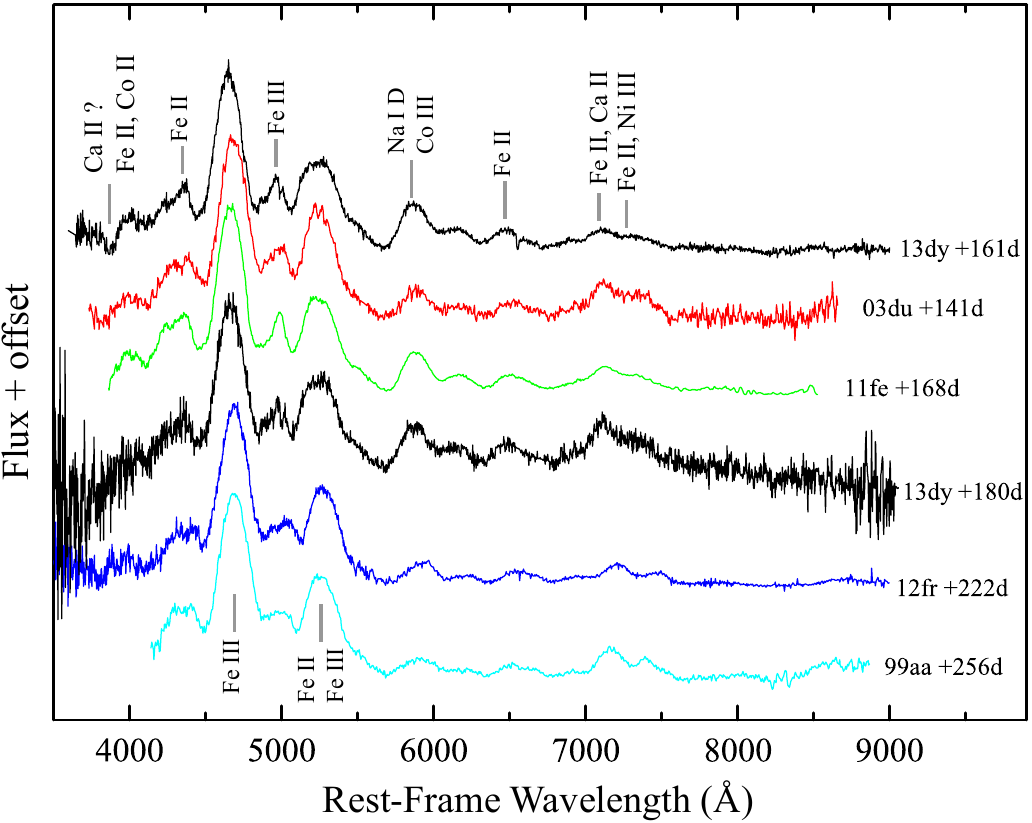}
 \caption{Spectra of SN 2013dy, SN 1999aa, SN 2003du, SN 2011fe and SN 2012fr \citep{Childress15} near the nebular phase. These spectra are normalized accordingly  based on the  region of 4400 -- 5000\AA. All of these spectra were corrected with redshift and reddening.}
\label{<SpNebu>}
\end{figure}

At $t > +4$ months,  as presented in Figure \ref{<SpNebu>}, all of these samples become to be uniform and are dominated by the emission of iron-group elements. However, the difference can be found in the strength of some features,  for example, the bump from 7000 to 7400\AA~, which can be attributed  to the emission of  [\FeII] $\lambda$7155 and [\ion{Ni}{2}] $\lambda$7378 at  $v\sim$ 2000 km s$^{-1}$. Such a bump might indicate that SN 2013dy is approaching the nebular phase. 

In summary, the above comparisons suggest that SN 2013dy shares the major similarities with  the normal SN Ia 2003du and SN 2011fe. On the other hand, the spectra of SN 2013dy are also very similar to those of SN 1999aa and iPTF 14bdn from one week before the $B$ band maximum.

\subsection{Velocities of Ejecta}
\label{subsect:EjeV}

\begin{figure}
\centering
\includegraphics[width=8.5cm,angle=0]{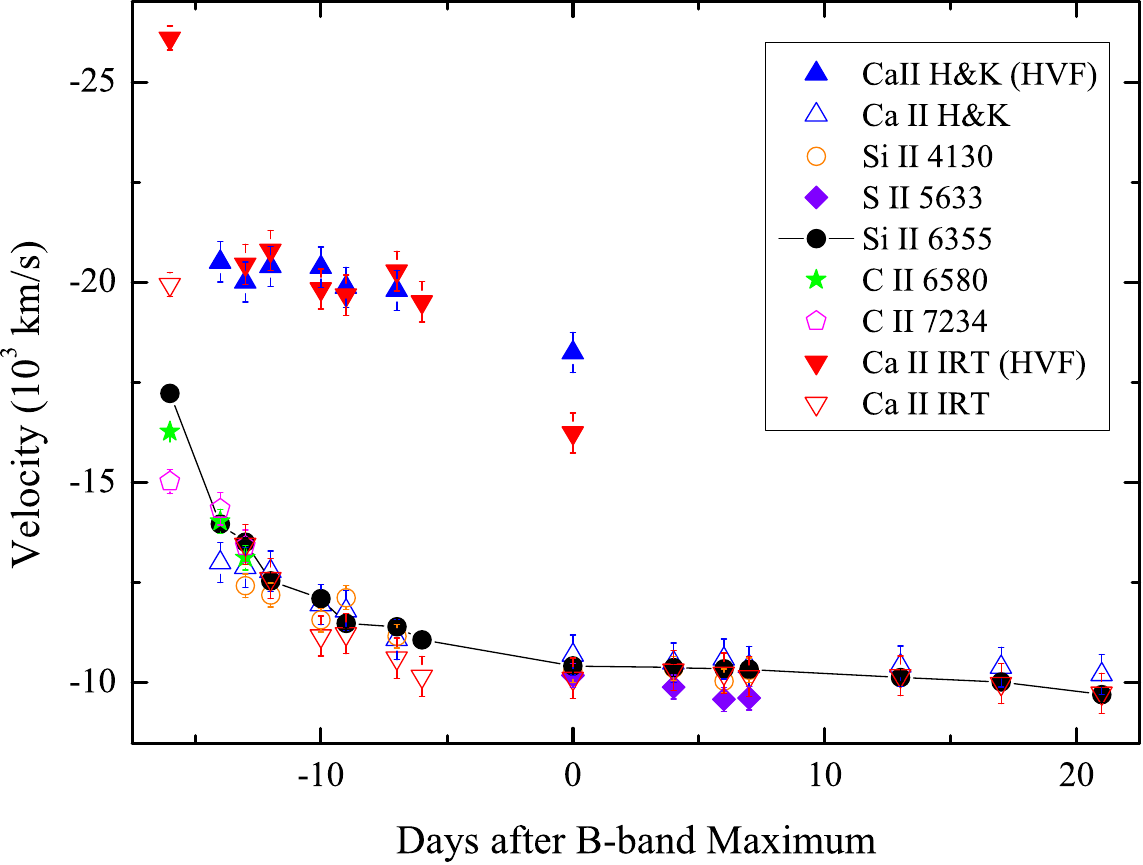}
 \caption{Velocity evolution of different elements inferred from the spectra of SN 2013dy.}
\label{<Vel>}
\end{figure}

Figure \ref{<Vel>} displays the ejecta velocities of SN 2013dy via the absorption features of some spectral lines, such as \CaII\,H$\&$K, \SiII\,$\lambda$4130, \ion{S}{2} $\lambda$5633, \SiII\,$\lambda$6355, \CII\,$\lambda$ 6580, \CII\,$\lambda$7234 and the \CaII\,IR triplet. The location of the absorption minimum was measured by using both the Gaussian fit routine and the direct measurement of the center of the absorption, and the results were averaged. 

The HVF of the \CaII\,IR triplet at $t\approx -16$ days is $\sim 26,000$ km s$^{-1}$ which is close to that of SN 2011fe at the same phase and  slower than that of SN 2012fr at $t\approx -14$ days (i.e., $\sim31,000 $km s$^{-1}$; \citealp{C13}). At this phase, the velocity of \SiII\,$\lambda$6355 is $\sim 17,200$ km s$^{-1}$, slower than that of the photospheric component of  \CaII\,IR triplet (i.e., $\sim$20,000 km s$^{-1}$) and faster than that of the \CII\,features (i.e., $\sim$16,300 km s$^{-1}$ and $\sim$15,000 km s$^{-1}$ for $\lambda$ 6580 and $\lambda$7234, respectively). Three days later, the velocity of \CII\,\ld6580 drops to about 13,000 \kms, which is close to the typical expansion velocity of this line in SNe Ia \citep{SilverCII}.   It is notable for the velocity plateaus of the HVFs of \CaII\,from $t\approx -14$ to $-6$ days at $v\approx 20,000$ km s$^{-1}$, while the velocities of photospheric components (e.g., \CaII\ and \SiII) are quickly declining.  

After the maximum light, the velocity of \SiII\,$\lambda$ 6355 is $v\sim$10,450 km s$^{-1}$ with a velocity gradient of $18\pm20$ km s$^{-1}$ day$^{-1}$. This gradient  is derived from the velocity from $t\approx$ +0 to +13 days and this line is not contaminated by the surrounding lines.  Such a low velocity gradient is similarly seen for the velocity evolution of \CaII, which  puts SN 2013dy into the LVG category of SNe Ia according to the classification scheme of \citet{Ben05}. On  the other hand, the velocity of \SII\,$\lambda$5633 is slower but declines quicker. The IMEs of SN 2013dy generally have similar expansion velocities (e.g., $\sim$10,000 km s$^{-1}$), suggestive of a relatively uniform distribution of the burning products in the ejecta.

\section{Discussion}
\label{sect:Disc}

\subsection{Distance}
\label{subsect:dist}

The observed velocity of NGC 7250 is 1166 km s$^{-1}$, which after correcting for the local group infall onto Virgo, Galaxy, and Sharply becomes 1410 km s$^{-1}$ \citep{Mould} or a distance D = 19.58 $\pm$ 2.0 Mpc  on the scale of H$_0$ = 72 km s$^{-1}$ Mpc$^{-1}$. On the other hand, the distance derived from the Tully-Fisher relation of this galaxy is D=13.7 $\pm$ 3.0 Mpc \citep{Tully,Nasonova11}, which was adopted in  Z13 and P15 as the distance of SN 2013dy. Furthermore, we can calculate the distance  from the WLR of SNe Ia \citep{Phillip93,Blondin12}. The brightness of SNe Ia in the $H$ band is relatively insensitive to the reddening and is more uniform around the peak compared to the corresponding values in the optical bands \citep{Meik00,Kri04,Barone-Nugent12}. Thus, we can estimate the distance to SN 2013dy based on its $H$ band light curve published in P15.

\begin{deluxetable}{ccc}
\tablewidth{0pt}
\tablecaption{The estimations for the distance of SN\,2013dy}

\tablehead{\colhead{Method} & \colhead{Details} & \colhead{Results (Mpc)} }
\startdata
Hubble Flow\tablenotemark{a}  & 1410 \kms & 19.6$\pm$2.0\\
Tully-Fisher\tablenotemark{b}  & $JHK$ bands & 13.7$\pm$3.0 \\
Phillips  Relation\tablenotemark{c}  & \DR = 0.90 & 17.0$\pm$3.0 \\
Phillips  Relation\tablenotemark{d}  & \DR = 0.90 & 19.5$\pm$3.0\\
NIR luminous\tablenotemark{e} & $H$ band & 31.3$\pm$3.0\tablenotemark{f}
\enddata
\tablenotetext{a}{Corrected for Virgo infall, GA and Shapley \citep{Mould}, on the scale of $H_0$=72 \kms Mpc$^{-1}$.}
\tablenotetext{b}{\citet{Tully} and \citet{Nasonova11}.}
\tablenotetext{c}{\DR = 0.90; \citet{Phillip93}.}
\tablenotetext{d}{Modified \citet{Phillip93} relation from the Fig. 13 of \citet{Blondin12} for the normal SNe Ia.}
\tablenotetext{e}{From the $H$ band light curve of SN 2013dy published in P15.}
\tablenotetext{f}{The error of extinction is involved.}
\label{Tab:disct}
\end{deluxetable}

Table \ref{Tab:disct} lists the distance derived from the above methods and an average value $D = 20.0\,\pm$\,4.0 Mpc is adopted. Note that the estimation from the $H$ band brightness (i.e., $D = 31.3\,\pm\,$3.0 Mpc) is much larger than that inferred from other methods. It suggests that the intrinsic $H$ band brightness of SN 2013dy  (i.e., $M(\rm{H}) = -17.33 \pm 0.30$ mag, if $D$ = 20.0 Mpc or $M(\rm{H}) = -16.51 \pm 0.24$ mag if $D$ = 13.7 Mpc )  is at least 1.0 mag lower than the normal (e.g., $M(\rm{H}) = -18.432\pm0.017$ mag, \citealp{Kattner12}; $-18.40\pm0.08$ mag, \citealp{Folatelli10};  $-18.30\pm0.04$ mag, \citealp{Barone-Nugent12}; $-18.314 \pm 0.024$ mag,  \citealp{Weyant14}). Besides, \citet{JJzhang15} pointed out that the peak $H$ band brightness of SN 2011hr (an extreme 91T-like event, \DR = 0.93 mag) is about 0.7 mag higher than the average. Therefore, the scatters of SNe Ia in $H$ band is about 2.0--2.5 mag for the SNe Ia with small decline rate (i.e., \DR\,$\approx$\,0.90). This larger scatter might be a challenge to the assumption that the NIR luminosities are more uniform in SNe Ia.

\subsection{Spectral Energy Distribution}
\label{subsect:SED}
\begin{figure}
\centering
\includegraphics[width=8.5cm,angle=0]{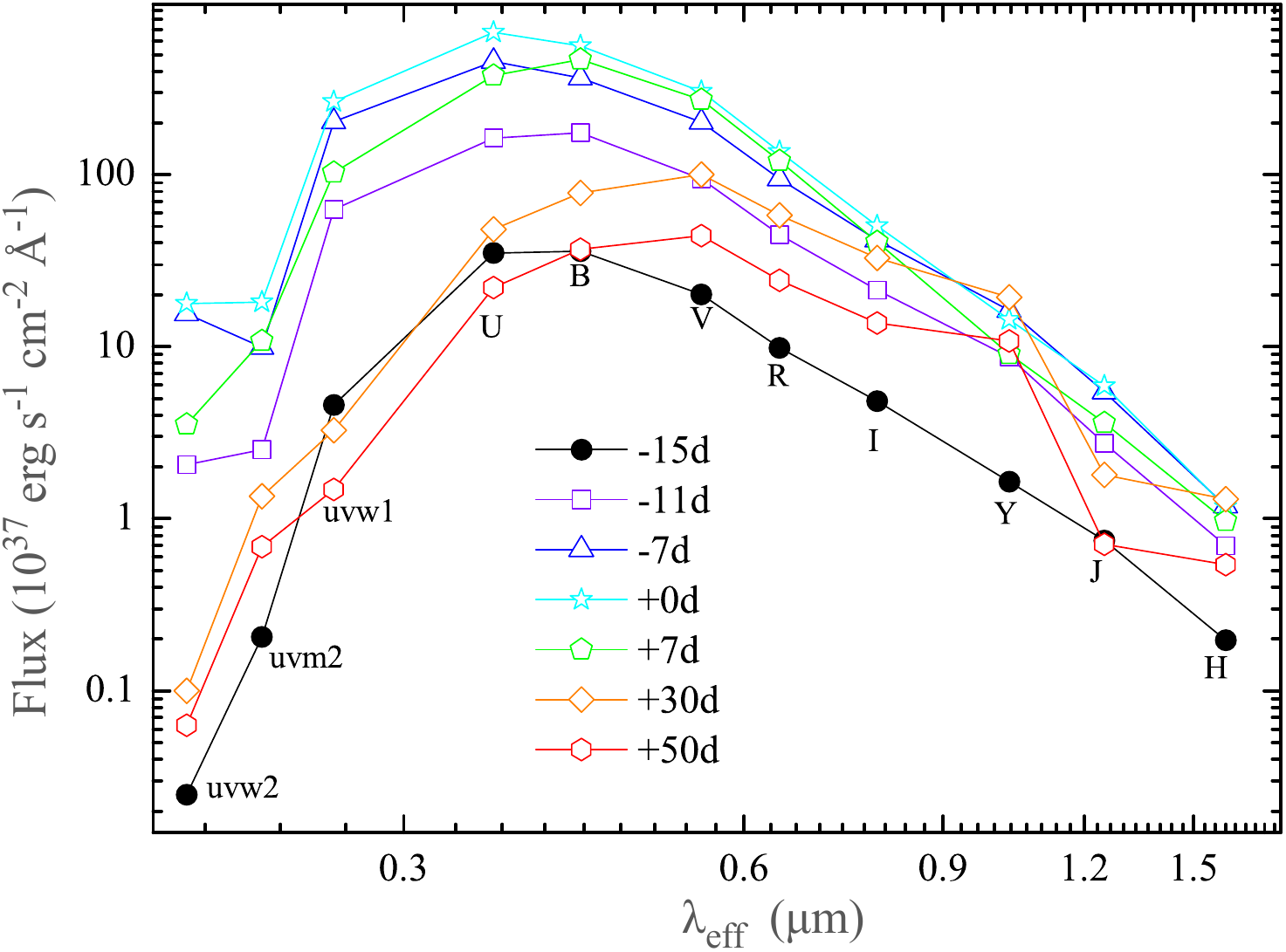}
 \caption{Spectral energy distribution of SN 2013dy at $t\approx-14, -11, -7, +0, +7, +30$, and +50 days. The positions of effective wavelength for each filter are marked.}
\label{<SED>}
\end{figure}

Based on the UV, optical, and NIR photometry (covering the wavelength from 1600 to 18,000\AA) presented in Figure \ref{<LC>}, we can construct the SED of SN 2013dy roughly through the observed fluxes in various passbands at the same epochs. The missing data can be obtained through interpolation of the adjacent data.  Figure \ref{<SED>} displays the SED of SN 2013dy at $t \approx -15, -11, -7, +0, +7, +30$, and +50 days. Note that the $uvw2$ and $uvw1$ fluxes are corrected for the ``red tail" effect \citep{redtail}.  This figure shows a clear energy translation of SN 2013dy from blue to red in wavelength.  A notable bump around the $Y$ band at $t \ge +30$ days might relate to the deficit in the $J$ and $H$ bands.

\begin{figure*}
\centering
\includegraphics[width=15cm,angle=0]{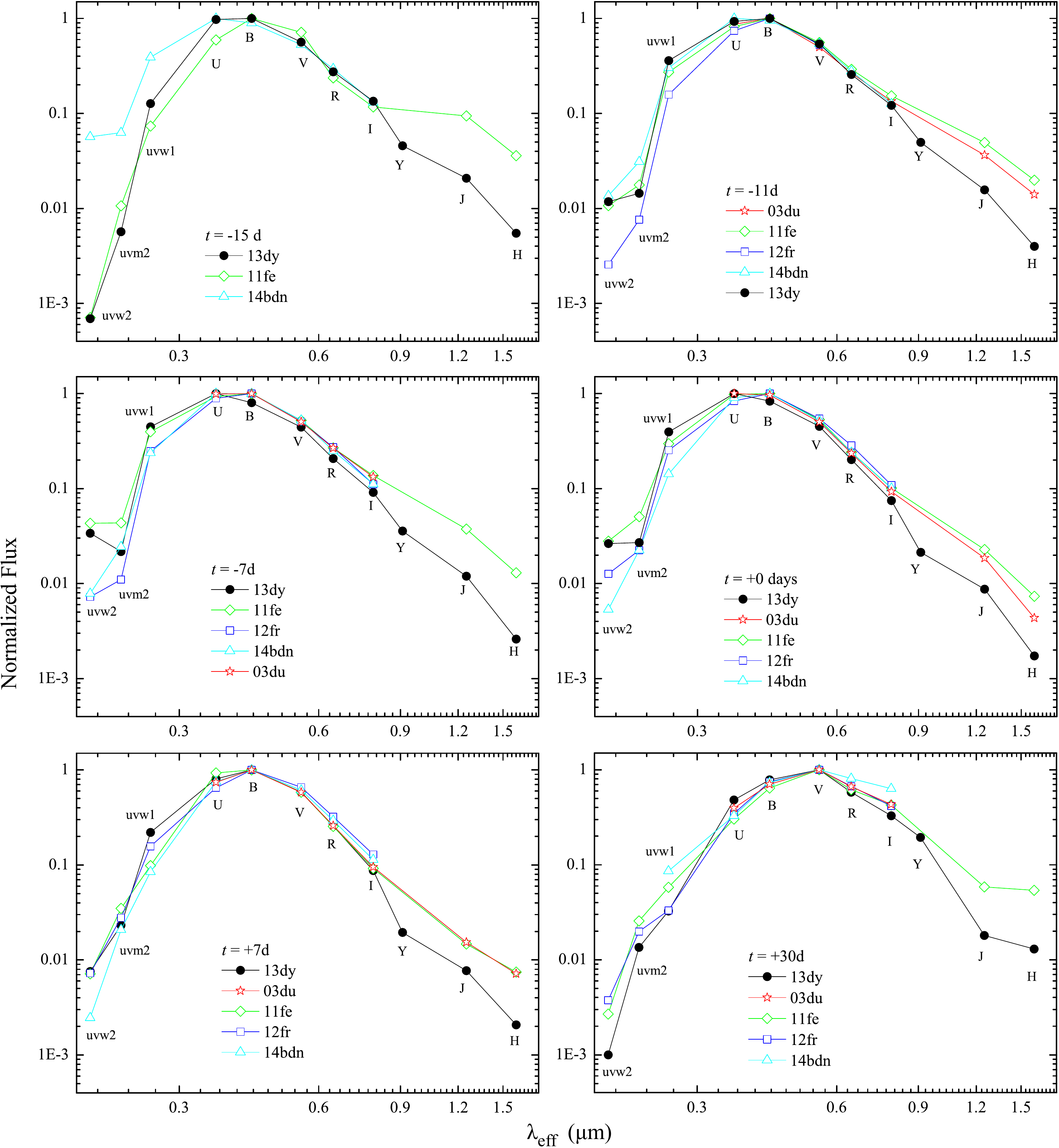}
 \caption{SED Comparison for SN 2003du, SN 2011fe, SN 2012fr, SN 2013dy, and iPTF 14bdn (99aa-like) at $t \approx -15, -11, -7, +0, +7$, and +30 days.}
\label{<SED_comp>}
\end{figure*}

To  understand the energy transmission of SN 2013dy, we compare  its SED with that of SN 2003du, SN 2011fe, SN 2012fr and iPTF 14bdn at six selected phases; see Figure \ref{<SED_comp>}. It is notable that the NIR fluxes of SN 2013dy are much lower than those in SN 2003du and SN 2011fe. The lower flux of SN 2013dy in the $J$ and $H$ bands conforms to the bluer $V-J$ and $V-H$ color compared to those in SN 2011fe, as reported in  fig. 5 of P15.

The SED could give a limit to the extinction of these SNe Ia. For example, the temperature of SN 2013dy should be lower than that of iPTF 14bdn at $t < -7$ days owing to the absence of \FeIII\,lines in the spectra of the former.  On the other hand, SN 2013dy might have a higher temperature than that of SN 2003du and SN 2011fe in light of  the comparison of spectral features. Furthermore, the reddenings of SN 2003du, SN 2011fe, and iPTF 14bdn are quite small (e.g., $E(B-V) <$ 0.03). The sequence of temperature could give a region of extinction, i.e., $0.20 \le E(B-V) \le 0.30$, which conforms to the estimation in Section \ref{subsect:ext}.

 \begin{figure*}
\centering
\includegraphics[width=13cm,angle=0]{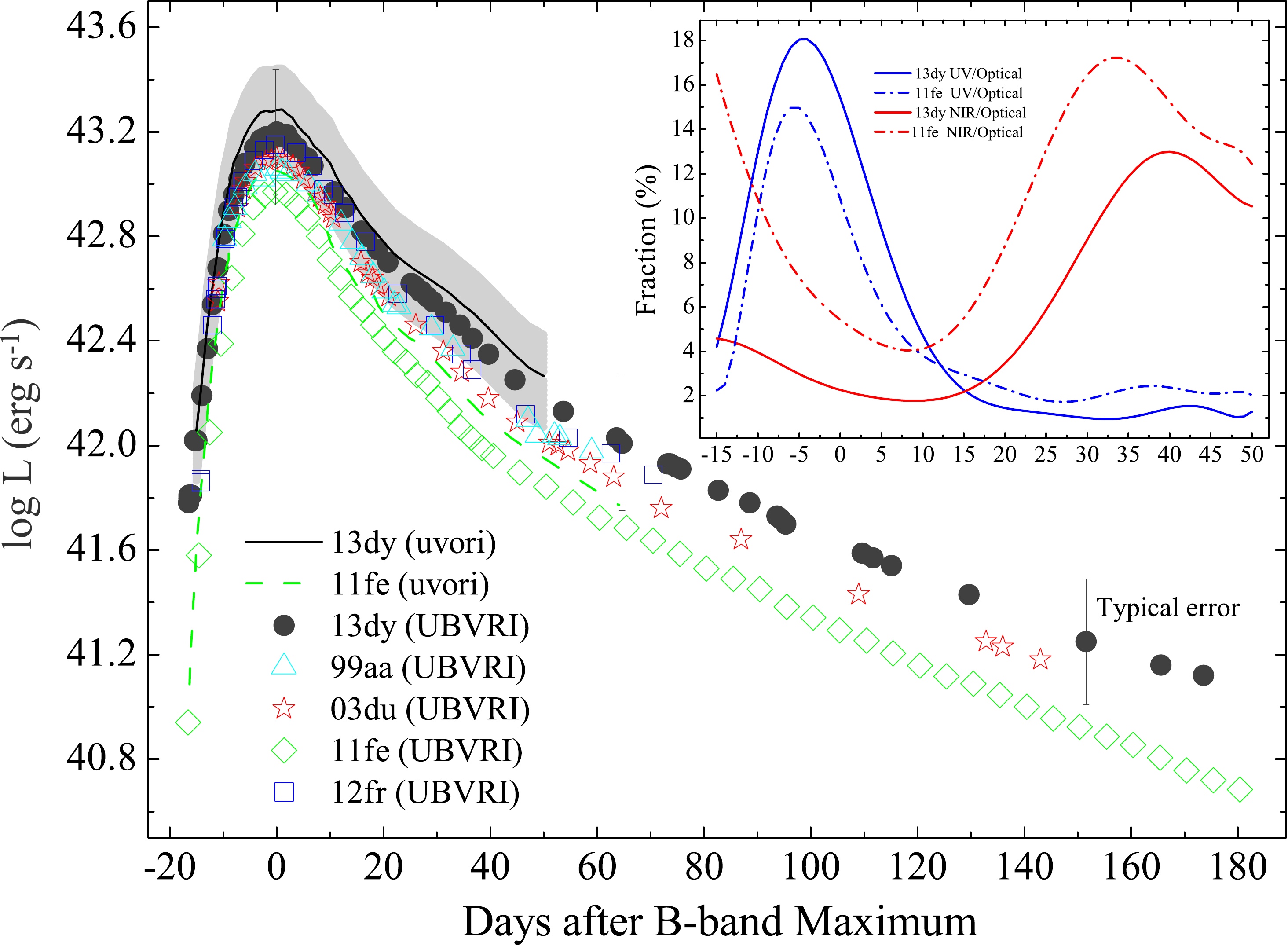}
 \caption{Quasi-bolometric $UBVRI$ (3300-9000\AA) light curves of SN 2013dy  compared with that of SN 1999aa, SN 2003du, SN 2011fe and SN 2012fr.  The `$uvoir$' (1600--24,000\AA) of SN 2013dy and SN 2011fe are overplotted. The grey area shows the 1$\sigma$ range of the `$uvoir$' bolometric luminosities of SN 2013dy considering the uncertainties from the distance (i.e., D = 20.0 $\pm$ 4.0 Mpc). The fraction of UV (1600--3300\AA) and NIR (9000--24,000\AA) flux to the optical flux for SN 2013dy and SN 2011fe are presented in the top right panel.}
\label{<bolo>}
\end{figure*}

\subsection{Bolometric Light Curve and the Mass of \Nifs}
\label{subsect:bolo}

Figure \ref{<bolo>} displays the quasi-bolometric  light curves of SN 2013dy derived from $UBVRI$ photometry, compared with that of SN 1999aa, SN 2003du, SN 2011fe, and SN 2012fr. SN 2013dy is similar to SN 2012fr but with slightly higher luminosity.

We calculate the UV-optical-NIR (``uvoir", covering the wavelength from 1600 to 24,000 \AA) bolometric light curves of SN 2013dy at $-15 < t < +50$ days based on the photometry presented in Figure \ref{<LC>}. Note that the UV flux  at $t < -10$ days is estimated through the stretched UV light curves of SN 2011fe,  which might induce some errors because of the difference presented in Figure \ref{<UVOP>}.  The fluxes beyond the $H$ band (i.e., 18,000 to 24,000\AA) are extrapolated through blackbody approximation. The missing fluxes at wavelengths shorter than the $Swift$ UV filters (e.g., $<$ 1600 \AA) or longer than 24,000 \AA~ are negligible. The ``uvoir" bolometric light curve of SN 2011fe \citep{KCzhang15} is also overplotted in this figure.  An obvious shoulder at around $t \approx +40$ days of SN 2013dy implies a higher intrinsic luminosity than SN 2011fe.

The fraction of the UV (1600-3300\AA) and NIR (9000-24,000\AA) to the optical flux of SN 2013dy  and SN 2011fe is presented in the top right panel of Figure \ref{<bolo>}.  Based on this panel, we can find out that the UV flux of SN 2013dy is relatively higher than that of SN 2011fe at the early phase while the NIR flux of the former is relatively lower. It also indicates that SN 2013dy has relatively stronger optical flux at $t\,>\,\sim\,+10$ days. 
"

Based on the $``uvoir"$ light curve, we estimate that SN 2013dy reaches its bolometric maximum ($L_{\rm max}$ = [1.95 $\pm$ 0.55]$\times$ 10$^{43}$ erg s$^{-1}$) at about 0.9 days before the $B$ band maximum. This could also be found in some bight SNe Ia (e.g., 91T/99aa-like events) and relates to the strong contribution at wavelengths shorter than the $B$ band.  The uncertainty of the peak flux includes the errors in the distance modulus, the observed magnitudes, the NIR corrections, and the missing of flux. Z13 found that the first-light time of SN 2013dy is JD 2456483.18, thus the rise time of the bolometric light curve (i.e., $t_{\rm rise}\approx$ 17.3 days) is adopted in the following estimation.  With the derived bolometric luminosity and the rise time of the bolometric light curve, the synthesized $^{56}$Ni mass estimated using the Arnett law \citep{Arn82,56Ni05}  is $M(^{56} {\rm Ni})=0.90\,\pm\,0.26\,{\rm M}_{\sun}$. This value is similar to that of SN 2012fr (0.88 M$_{\sun}$, \citealp{JJzhang14}) and  larger than that of SN 1999aa (0.72  M$_{\sun}$, this paper), SN 2003du (i.e., 0.63\,$\pm$\,0.19 M$_{\sun}$, this paper; 0.68 M$_{\sun}$, \citealp{03du}; 0.60 M$_{\sun}$, \citealp{Stritzinger06}) and SN 2011fe (e.g., 0.53 M$_{\sun}$, \citealp{Munari13}; 0.56 M$_{\sun}$, \citealp{Mazzali14}).

On the other hand, we could estimate some explosion parameters of this SN by adopting the same method as in \citet{Tau09dc}. The ejecta mass ($M_{\rm{ej}}$) and the total explosion energy  ($E_{\rm{ej}}$) can be derived from the diffuse time $\tau_{\rm m}$ and ejecta velocity $v$, where $M_{\rm ej} \propto \tau^2_{\rm m} v$ and  $E_{\rm kin} \propto \tau^2_{\rm m} v^3$.  The diffuse time of SN 2013dy is 27.58 days, which is longer than that of SN 2011fe (i.e., 25.33 days). On the other hand, the velocity of \SiII\,\ld6355 of SN 2013dy at around maximum is 10,450 km/s which is also slightly faster than that of SN 2011fe (i.e., 10,340 km/s).   As a result, we find that $M_{\rm{ej, 13dy}}\approx1.20M_{\rm{ej,11fe}} $ and $E_{\rm{ej, 13dy}}\approx1.22E_{\rm{ej,11fe}}$. Note that these coefficients conform to the stretch factors adopted in Figure \ref{<LC>} for the light curve comparison.  The mass of $^{56}$Ni produced in SN 2013dy is 0.67 M$_{\sun}$ based on the estimation of SN 2011fe (e.g., 0.56M$_{\sun}$, \citealp{Mazzali14}) and the similar component assumption in Arnett's (1982) law. This result is independent of distance and extinction. It is smaller than the estimation of the bolometric curve. A smaller estimation for the \Nifs\ mass of SN 2013dy might suggest a distinct energy and material distribution between SN 2013dy and SN 2011fe since only the opaque mass can be calculated from these simple analytic approaches.

\begin{figure*}
\centering
\includegraphics[width=16cm,angle=0]{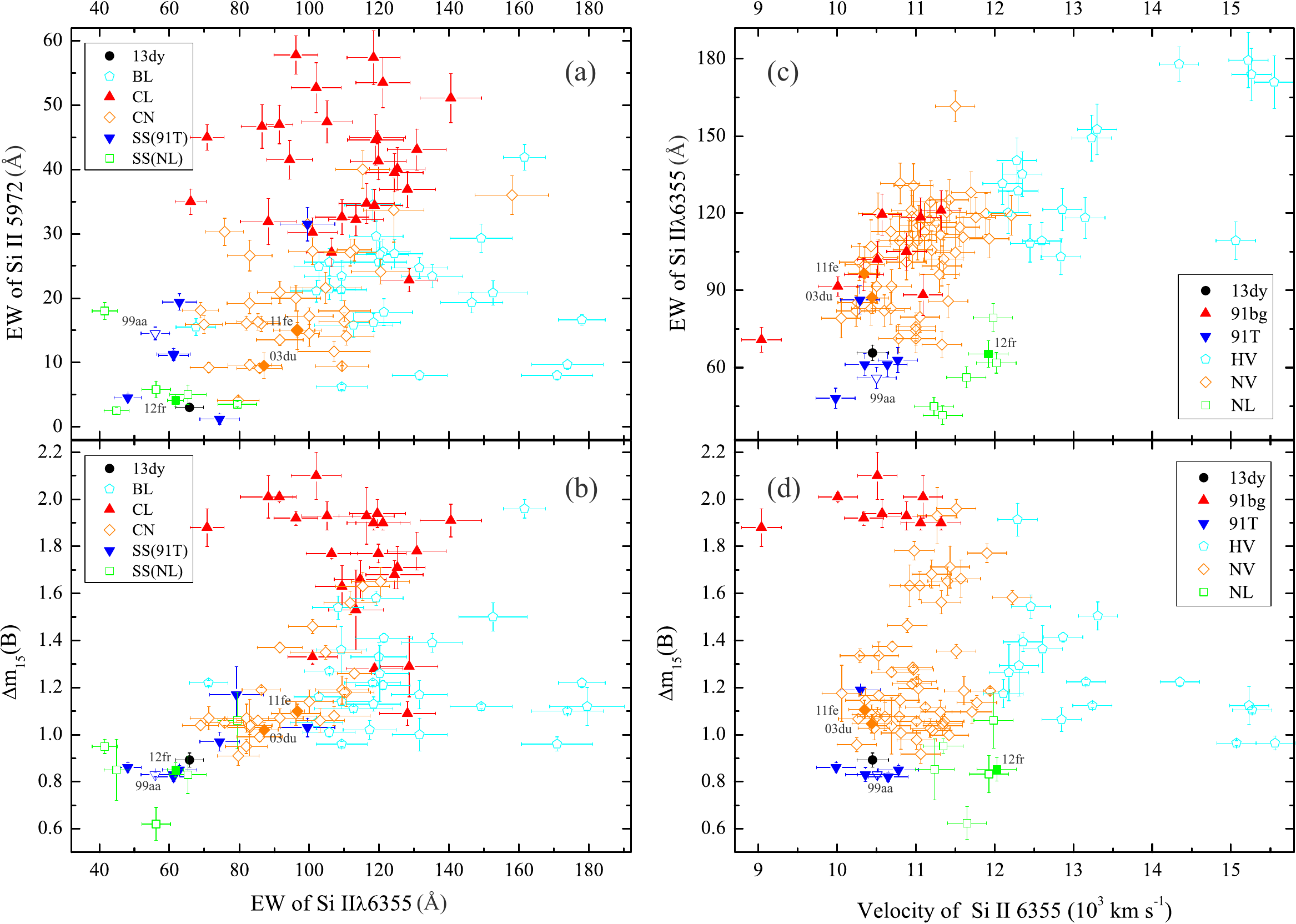}
 \caption{Comparison of various spectroscopic and photometric indicators from SN 2013dy with those from other SNe Ia as measured by \citet{Blondin12}, \citet{Silver12}, \citet{Wang09a}, \citet{JJzhang14} and this paper. The selected sample have spectra within $t\pm3$ days. Panel (a) and (b):  the EW of \SiII\,$\lambda$5972 and $\Delta m _{15}$(B)  vs. the EW of \SiII\,$\lambda$6355 at maximum light with subclasses defined by \citet{Branch09};  panel (c) and (d): the EW of \SiII\,$\lambda$6355 and  $\Delta m _{15}$(B)  vs. the velocity of \SiII\,6355 at maximum light with subclasses defined by \citet{Wang09a}.}
\label{<diver>}
\end{figure*}

\subsection{Spectroscopy Classification}
\label{subsect:diver}
Spectroscopy classification schemes have recently been proposed to highlight the diversity of relatively normal SNe Ia. \citet{Ben05} classified the normal SNe Ia into LVG and HVG groups by the temporal velocity gradient of the \SiII\,$\lambda$6355 line.  Based on the EW of the absorption features of \SiII\,$\lambda$5972 and \SiII\,$\lambda$6355,  \citet{Branch06,Branch09} suggested dividing the SN Ia sample into four groups: cool (CL), shallow silicon (SS), core normal (CN), and broad line (BL). The CL group  mainly consists of the faint objects like SN 1991bg \citep{Filip92b}. Besides of the faint peculiar events like SN 2002cx \citep{Li02cx}, the SS group is mainly consisted by the bright 91T/99aa-like events, the superluminous events (e.g., SN 2003fg, \citealp{Howell06}; SN 2007if, \citealp{Scalzo07if} and SN 2009dc, \citealp{Tau09dc,Silver11}) and also the narrow-lined SNe Ia (NL; e.g., SN 2012fr; \citealp{JJzhang14}).  We  note that the small EW of   \SiII\,$\lambda$6355 could be derived from different profiles of this line, for example, the absorption with smaller depth and larger width (e.g., 91T/99aa-like events) or the other with larger depth and small width (e.g., NL). Therefore, we divide the bright but generally normal members of SS into two subgroups, SS/91T-like and SS/NL, in the following discussion. On the other hand, \citet{Wang09a} proposed using the expansion velocity of the \SiII\,$\lambda$6355 line to distinguish the subclass with a higher \SiII\,velocity (HV) from that with a normal velocity (NV).  The HV SNe Ia are found to have redder $\bv$ colors \citep{Wang09a} and different locations within host galaxies in comparison to the NV ones \citep{Wang13}, suggesting that the properties of their progenitors may be different.

The spectroscopy classification of SN 2013dy is compared with the  large SNe Ia sample  from the spectral spectral data sets of CfA \citep{Blondin12} and the Berkeley SuperNova Ia Program \citep{Silver12} at $t\approx \pm$ 3 days. However, SN 2013dy resembles the normal SN  2003du and SN 2011fe observationally.  The weak \SiII\,absorptions (i.e., $\sim$70.7\AA ~ and  $\sim$3.3\AA ~ for  \SiII\,$\lambda$6355 and $\lambda$5972, respectively), however,  put it into the transitional region of  SS and CN at the near-side of the former in the \citet{Branch06} diagram, as shown in Figure \ref{<diver>} (a) and (c). It is difficult to distinguish the SS/91T-like from SS/NL in this scheme.  SN 2013dy is also close to the NL SN 2012fr.  However, the 91T/99aa-like events keep away from NL SNe Ia  in the \citet{Wang09a} diagram owing to the difference in velocity,  as shown in Figure \ref{<diver>} (b) and (d). In the  \citet{Wang09a}  scheme,  SN 2013dy and SN 2012fr are also separated and the former resides at the border between the NV and the 91T/99aa-like subclasses.  The slow decline rate, the small EW of \SiII\,absorptions, and the low velocity gradient resemble the properties of the 91T/99aa-like objects.  Besides, the $U-B$ color of SN 2013dy is as blue as that of SN 1999aa at the early phase.  However, the absence of \FeIII\, lines in the early spectra of SN 2013dy is a clear distinction from 91T/99aa-like events. 

\section{Conclusion}
\label{sect:con}
We have presented extended observations of Type Ia SN 2013dy obtained at LJT, XLT, TNT and $Swift$-UOVT.  Combined with the data published in Z13 and P15, this target is a well-sampled SN Ia with wide wavelength coverage (i.e., from $\sim$1600\AA~ to $\sim$18,000\AA). In general, this SN resembles the normal SNe Ia (e.g., SN 2003du and SN 2011fe) in both photometry and spectroscopy. Nevertheless, it also shares some similarities with the 99aa-like events, such as the $U-B$ and ``$uvm2-uu$"  color before maximum, the small decline rate, the low velocity gradient, and the small ratios of $R$(\SiII), $R$(S/Fe) and $R$(Si/Fe).  However, the absence of \FeIII\ lines in the early spectra of SN 2013dy might exclude it from 91T/99aa-like group. 

A problem is the uncertainty of the distance. In general, the distance (i.e., D $\approx$ 20.0 Mpc) derived from several methods seems to be more reasonable than the estimations from the Tully-Fisher relation of the host galaxy (i.e., D $\approx$ 13.7 Mpc). Based on the larger one, we estimated the peak brightness (e.g., $M_{\rm max}$(B) = $-19.65 \pm 0.40$ mag), maximum bolometric luminosity ([$1.95\pm 0.55] \times 10^{43}$ erg s$^{-1}$) and synthesized nickel mass (0.90 $\pm$ 0.26 M$_{\sun}$). The distance derived from Tully-Fisher relation will yield fainter results (i.e.,   $M_{\rm B} = -18.83$ $\pm$ 0.3 mag; $L_{\rm max} = [9.05 \pm 2.50] \times10^{42}$ erg s$^{-1}$; $M$($^{56}$Ni) = 0.42 $\pm$ 0.15M$_{\sun}$) that are dimmer than the average of SNe Ia and do not  follow the WLR of SNe Ia well.  Additionally, the $H$ band luminosity of SN 2013dy is at least 1.0 mag fainter than the average of SNe Ia. That might indicate inefficient emission translation from optical to NIR for SN 2013dy or imply that this SN is an intrinsically fainter SNe Ia, although it shows a small \DR. Moreover, the $H$ band photometry of SN 2013dy and SN 2011hr suggests a larger  scatter (i.e., $\sim$2.0--2.5 mag) among the SNe Ia with a slow decline rate (e.g., \DR\,$\approx$ 0.90 mag) and might challenge the assumption of the uniform NIR luminosity for SNe Ia.

On the other hand, it is difficult to put SN 2013dy into one certain subclass in the current classification system. The spectroscopy classification reveals that SN 2013dy might be a transitional event  residing on the border of different subclasses, such as SS-CN in the \citet{Branch06} classification scheme or NV-91T/99aa-like in the  \citet{Wang09a} classification system. Moreover, the diversity of the SNe Ia with small \DR and EWs of \SiII\,  (e.g., the superluminous  SNe Ia, the 91T/99aa-like events, the SN 2012fr-like events, and SN 2013dy) might imply the only EW criterion is insufficient  for the classification of this sample and indicate a complex construction of their ejecta.

All of these suggest that SN 2013dy might not be a typical normal SNe Ia, although it is normal in some respects.  Further modeling work is essential to reveal the nature of this SN.

\acknowledgments
We thank the anonymous referee for his/her constructive suggestions that helped to improve the paper. We acknowledge the support of the staff of the Li-Jiang 2.4 m telescope (LJT), Xin-Long 2.16 m telescope, and Tsinghua-NAOC 0.8 m telescope (TNT). Funding for the LJT has been provided by Chinese academe of science (CAS) and the People's Government of Yunnan Province.   The TNT is owned by Tsinghua University and operated by the National Astronomical Observatory of the Chinese Academy of Sciences (NAOC).  The data from UVOT comes from the $Swift$ Data Center. And we also thank Dr. Wei-Kang Zheng of UC Berkeley who provided the early $BVRI$  photometry of SN 2013dy (Z13).

Financial support for this work has been provided by the National Science Foundation of China (NSFC, grants 11403096, 11178003, 11325313, 11133006, 11361140347, 11203034, 11203078, 11573069, 11303085, 11203070); the Major State Basic Research Development Program (2013CB834903); the Strategic Priority Research Program ``The Emergence of Cosmological Structures" of the CAS (grant No. XDB09000000); the Key Research Program of the CAS (Grant NO. KJZD-EW-M06);  the CAS ``Light of West China" Program; the Youth Innovation Promotion Association of the CAS; the Open Project Program of the Key Laboratory of Optical Astronomy, NAOC, CAS; and the key Laboratory for Research in Galaxies and Cosmology of the CAS.

\begin{deluxetable*}{cccccccc}
\tablenum{2}
\tablewidth{0pt}
\tablecaption{$Continue$}
\tablehead{\colhead{MJD} & \colhead{Day\tablenotemark{a}} & \colhead{$U$(mag)} & \colhead{$B$(mag)} & \colhead{$V$(mag)} & \colhead{$R$(mag)} & \colhead{$I$(mag)} & \colhead{Telescope} }

\startdata
56529.05	&	28.17	&	15.60(03)	&	15.31(01)	&	14.10(01)	&	13.73(01)	&	13.45(01)	&	TNT	\\
56530.03	&	29.15	&	15.71(05)	&	15.40(02)	&	14.14(01)	&	13.73(01)	&	13.44(01)	&	TNT	\\
56532.49	&	31.61	&	15.84(04)	&	15.56(03)	&	14.28(02)	&	13.81(02)	&	13.43(03)	&	LJT	\\
56535.21	&	34.33	&	16.01(03)	&	15.76(01)	&	14.42(01)	&	13.94(01)	&	13.49(01)	&	TNT	\\
56537.50	&	36.62	&	16.05(02)	&	15.90(01)	&	14.55(01)	&	14.05(01)	&	13.58(01)	&	LJT	\\
56540.51	&	39.63	&	16.23(04)	&	16.04(02)	&	14.68(03)	&	14.23(03)	&	13.74(02)	&	LJT	\\
56545.50	&	44.62	&	16.45(03)	&	16.20(02)	&	14.89(02)	&	14.50(02)	&	14.09(03)	&	LJT	\\
56546.34	&	45.46	&	\nodata	&	16.14(02)	&	14.93(01)	&	14.55(01)	&	14.14(01)	&	TNT	\\
56547.34	&	46.46	&	\nodata	&	16.21(03)	&	14.97(01)	&	14.61(01)	&	14.23(01)	&	TNT	\\
56550.32	&	49.44	&	\nodata	&	16.24(02)	&	15.05(01)	&	14.68(01)	&	14.37(01)	&	TNT	\\
56554.52	&	53.64	&	16.68(04)	&	16.35(03)	&	15.18(03)	&	14.88(02)	&	14.55(02)	&	LJT	\\
56560.32	&	59.44	&	\nodata	&	16.41(02)	&	15.33(01)	&	15.03(01)	&	14.81(01)	&	TNT	\\
56561.31	&	60.43	&	\nodata	&	16.43(02)	&	15.37(01)	&	15.06(01)	&	14.88(01)	&	TNT	\\
56562.30	&	61.42	&	\nodata	&	16.41(02)	&	15.38(01)	&	15.10(01)	&	14.93(01)	&	TNT	\\
56563.30	&	62.42	&	\nodata	&	16.51(04)	&	15.41(02)	&	15.19(01)	&	14.95(01)	&	TNT	\\
56564.50	&	63.62	&	16.94(03)	&	16.44(01)	&	15.47(01)	&	15.18(01)	&	15.01(01)	&	LJT	\\
56565.50	&	64.62	&	16.98(03)	&	16.54(01)	&	15.48(01)	&	15.21(01)	&	15.08(01)	&	LJT	\\
56574.05	&	73.17	&	17.21(13)	&	16.56(03)	&	15.69(02)	&	15.49(02)	&	15.41(02)	&	TNT	\\
56574.53	&	73.65	&	17.21(03)	&	16.62(01)	&	15.67(01)	&	15.46(01)	&	15.41(01)	&	LJT	\\
56575.52	&	74.64	&	17.24(03)	&	16.65(01)	&	15.71(01)	&	15.49(01)	&	15.46(01)	&	LJT	\\
56576.10	&	75.22	&	\nodata	&	16.58(02)	&	15.73(01)	&	15.49(01)	&	15.46(01)	&	TNT	\\
56576.52	&	75.64	&	17.27(03)	&	16.67(01)	&	15.74(01)	&	15.52(01)	&	15.50(01)	&	LJT	\\
56576.96	&	76.08	&	\nodata	&	16.66(02)	&	15.80(01)	&	15.56(01)	&	15.55(01)	&	TNT	\\
56579.97	&	79.09	&	\nodata	&	16.73(02)	&	15.83(01)	&	15.62(01)	&	15.67(01)	&	TNT	\\
56583.52	&	82.64	&	17.43(03)	&	16.80(01)	&	15.93(01)	&	15.71(01)	&	15.77(01)	&	LJT	\\
56584.96	&	84.08	&	\nodata	&	16.68(03)	&	16.00(02)	&	15.76(02)	&	15.82(02)	&	TNT	\\
56585.98	&	85.10	&	\nodata	&	16.72(03)	&	15.96(02)	&	15.83(01)	&	15.86(02)	&	TNT	\\
56586.97	&	86.09	&	17.48(13)	&	16.84(03)	&	16.01(02)	&	15.87(02)	&	15.90(02)	&	TNT	\\
56588.95	&	88.07	&	\nodata	&	16.82(02)	&	16.08(01)	&	15.90(01)	&	16.02(01)	&	TNT	\\
56589.52	&	88.64	&	17.58(03)	&	16.86(01)	&	16.08(01)	&	15.91(01)	&	15.99(01)	&	LJT	\\
56589.98	&	89.10	&	\nodata	&	16.85(02)	&	16.11(01)	&	15.96(01)	&	16.06(01)	&	TNT	\\
56594.52	&	93.64	&	17.73(03)	&	16.92(01)	&	16.21(01)	&	16.08(01)	&	16.17(01)	&	LJT	\\
56595.18	&	94.30	&	17.76(03)	&	16.95(01)	&	16.25(01)	&	16.11(01)	&	16.20(01)	&	LJT	\\
56596.18	&	95.30	&	17.82(03)	&	16.99(01)	&	16.28(01)	&	16.14(01)	&	16.23(01)	&	LJT	\\
56601.96	&	101.08	&	\nodata	&	17.03(02)	&	16.39(01)	&	16.33(01)	&	16.46(01)	&	TNT	\\
56602.99	&	102.11	&	\nodata	&	17.07(02)	&	16.41(01)	&	16.33(01)	&	16.50(01)	&	TNT	\\
56603.98	&	103.10	&	\nodata	&	17.07(02)	&	16.40(01)	&	16.34(01)	&	16.46(01)	&	TNT	\\
56606.96	&	106.08	&	\nodata	&	17.12(02)	&	16.38(01)	&	16.37(01)	&	16.51(01)	&	TNT	\\
56608.00	&	107.12	&	\nodata	&	17.16(02)	&	16.52(01)	&	16.47(01)	&	16.68(01)	&	TNT	\\
56608.99	&	108.11	&	\nodata	&	17.15(02)	&	16.52(01)	&	16.52(01)	&	16.72(02)	&	TNT	\\
56610.49	&	109.61	&	18.01(03)	&	17.19(01)	&	16.52(01)	&	16.57(01)	&	16.65(01)	&	LJT	\\
56610.99	&	110.11	&	\nodata	&	17.18(02)	&	16.51(01)	&	16.49(01)	&	16.74(01)	&	TNT	\\
56612.01	&	111.13	&	\nodata	&	17.25(02)	&	16.69(02)	&	16.59(02)	&	16.80(02)	&	TNT	\\
56612.48	&	111.60	&	18.23(03)	&	17.23(01)	&	16.61(01)	&	16.62(01)	&	16.72(01)	&	LJT	\\
56612.96	&	112.08	&	\nodata	&	17.22(02)	&	16.57(02)	&	16.60(02)	&	16.81(02)	&	TNT	\\
56614.01	&	113.13	&	\nodata	&	17.19(03)	&	16.53(02)	&	16.47(02)	&	16.83(02)	&	TNT	\\
56615.12	&	114.24	&	\nodata	&	17.21(03)	&	16.67(02)	&	16.54(02)	&	16.84(02)	&	TNT	\\
56615.98	&	115.10	&	18.26(13)	&	17.28(02)	&	16.67(02)	&	16.69(02)	&	16.86(02)	&	TNT	\\
56616.98	&	116.10	&	\nodata	&	17.27(03)	&	16.56(05)	&	16.74(01)	&	16.94(02)	&	TNT	\\
56617.97	&	117.09	&	\nodata	&	17.33(02)	&	16.74(01)	&	16.79(01)	&	16.96(02)	&	TNT	\\
56618.98	&	118.10	&	\nodata	&	17.33(02)	&	16.78(02)	&	16.80(01)	&	17.03(02)	&	TNT	\\
56619.97	&	119.09	&	\nodata	&	17.35(03)	&	16.80(02)	&	16.86(02)	&	17.08(02)	&	TNT	\\
56620.98	&	120.10	&	\nodata	&	17.29(02)	&	16.77(01)	&	16.87(01)	&	17.08(02)	&	TNT	\\
56621.98	&	121.10	&	\nodata	&	17.31(02)	&	16.80(01)	&	16.88(01)	&	17.09(02)	&	TNT	\\
56628.47	&	127.59	&	\nodata	&	17.45(01)	&	16.92(01)	&	17.06(01)	&	17.17(01)	&	LJT	\\
56629.96	&	129.08	&	\nodata	&	17.46(03)	&	16.97(02)	&	17.09(02)	&	17.36(02)	&	TNT	\\
56630.49	&	129.61	&	18.48(05)	&	17.48(01)	&	17.00(01)	&	17.11(01)	&	17.20(01)	&	LJT	\\
56631.95	&	131.07	&	\nodata	&	17.50(02)	&	17.00(01)	&	17.15(01)	&	17.32(02)	&	TNT	\\
56632.98	&	132.10	&	\nodata	&	17.56(02)	&	17.08(01)	&	17.22(02)	&	17.39(02)	&	TNT	\\
56633.96	&	133.08	&	\nodata	&	\nodata	&	16.97(02)	&	17.20(02)	&	17.48(02)	&	TNT	\\
56635.96	&	135.08	&	\nodata	&	17.52(02)	&	17.10(02)	&	17.24(01)	&	17.44(02)	&	TNT	\\
56636.92	&	136.04	&	\nodata	&	\nodata	&	16.94(02)	&	17.17(02)	&	17.57(02)	&	TNT	\\
56639.93	&	139.05	&	\nodata	&	17.64(03)	&	17.21(02)	&	17.34(02)	&	17.55(02)	&	TNT	\\
56641.95	&	141.07	&	\nodata	&	17.62(03)	&	17.18(02)	&	17.42(02)	&	17.51(02)	&	TNT	\\
56645.96	&	145.08	&	\nodata	&	17.72(02)	&	17.20(02)	&	17.40(02)	&	\nodata	&	TNT	\\
56646.96	&	146.08	&	\nodata	&	17.69(02)	&	17.18(02)	&	17.49(02)	&	\nodata	&	TNT	\\
56647.54	&	146.66	&	\nodata	&	17.69(02)	&	17.22(02)	&	17.54(01)	&	17.67(04)	&	LJT	\\
56647.96	&	147.08	&	\nodata	&	17.77(03)	&	17.28(02)	&	17.61(03)	&	\nodata	&	TNT	\\
56648.95	&	148.07	&	\nodata	&	17.76(03)	&	17.28(02)	&	17.48(02)	&	\nodata	&	TNT	\\
56649.95	&	149.07	&	\nodata	&	17.78(03)	&	17.34(02)	&	17.65(03)	&	\nodata	&	TNT	\\
56652.49	&	151.61	&	19.02(08)	&	17.87(02)	&	17.39(01)	&	17.63(03)	&	17.76(03)	&	LJT	\\
56661.53	&	160.65	&	\nodata	&	17.98(02)	&	17.52(02)	&	17.76(02)	&	17.96(03)	&	LJT	\\
56664.55	&	163.67	&	\nodata	&	18.04(02)	&	17.61(02)	&	17.80(02)	&	17.97(04)	&	LJT\\
56666.49	&	165.61	&	19.28(08)	&	18.07(03)	&	17.64(01)	&	17.89(02)	&	17.99(04)	&	LJT	\\
56671.49	&	170.61	&	\nodata	&	18.14(02)	&	17.76(02)	&	18.01(02)	&	18.14(06)	&	LJT	\\
56674.50	&	173.62	&	19.47(09)	&	18.15(02)	&	17.71(01)	&	18.13(03)	&	18.21(05)	&	LJT	\\
56681.49	&	180.61	&	\nodata	&	18.24(03)	&	17.77(01)	&	18.17(02)	&	18.31(05)	&	LJT	\\
56683.49	&	182.61	&	\nodata	&	18.27(04)	&	17.79(02)	&	18.23(03)	&	18.35(06)	&	LJT		
\enddata

\end{deluxetable*}

\end{document}